\documentclass[lettersize,journal]{IEEEtran}
\usepackage{amsmath,amsfonts}
\usepackage{algorithmic}
\usepackage{algorithm}
\usepackage{array}
\usepackage[caption=false,font=normalsize,labelfont=sf,textfont=sf]{subfig}
\usepackage{textcomp}
\usepackage{stfloats}
\usepackage{url}
\usepackage{verbatim}
\usepackage{graphicx}
\usepackage{cite}
\usepackage{multirow}
\usepackage{arydshln}
\usepackage{hyperref}
\usepackage{cancel}
\usepackage{xcolor}

\begin{document}

\title{Embedding based Encoding Scheme for \\Privacy Preserving Record Linkage}


\author{Sirintra~Vaiwsri and Thilina~Ranbaduge
\thanks{Sirintra Vaiwsri is with the Faculty of Industrial Technology and Management, King Mongkut's University of Technology North Bangkok, Prachin Buri, 25230, Thailand. E-mail: sirintra.v@itm.kmutnb.ac.th \\
Thilina Ranbaduge is with Data61, Canberra,
ACT, 2600, Australia. E-mail: thilina.ranbaduge@data61.csiro.au}}



\maketitle

\begin{abstract}
To discover new insights from data, there is a growing need
to share information 
that is often held by different organisations. One key task in data
integration is the calculation of similarities between records in different databases to
identify pairs or sets of records that correspond to the same
real-world entities. Due to privacy and confidentiality concerns,
however, the owners of sensitive databases are often not allowed or
willing to exchange or share their data with other organisations to
allow such similarity calculations. Privacy-preserving record linkage 
(PPRL) is the process of matching records that refer to the same entity
across sensitive databases held by different organisations while ensuring 
no information about the entities is revealed to the participating 
parties. In this paper, we study how embedding based encoding techniques can be 
applied in the PPRL context to ensure the privacy of the entities that 
are being linked. We first convert individual q-grams into the embedded space
and then convert the embedding of a set of q-grams of a given record into 
a binary representation. The final binary representations can be used to link
records into matches and non-matches. We empirically evaluate our proposed 
encoding technique against different real-world datasets. The results
suggest that our proposed encoding approach can provide better linkage 
accuracy and protect the privacy of individuals against attack compared to state-of-the-art techniques for short record values. 
\end{abstract}

\begin{IEEEkeywords}
Q-gram embedding, Binarisation, Binary strings.
\end{IEEEkeywords}

\section{Introduction}
\label{introduction}

\IEEEPARstart{O}{rganisations} in many domains increasingly collect 
large databases
containing millions of records, where these records contain detailed
information about individuals, such as customers, patients, taxpayers,
or travellers. Often, such databases need to be shared and
integrated to facilitate advanced analytics and
processing~\cite{Chr12}.

Integrating databases can help to identify similar records that
correspond to the same real-world entities. Linked records allow 
improvement of data quality, enrichment of the information known
about individual entities, and facilitate the discovery of novel 
patterns and relationships between the entities that are 
represented by records in linked databases. However, this process 
is challenging because no unique entity identifiers, such as 
social security numbers, are available in the databases to be linked.
Therefore, quasi-identifying attributes such as names and addresses
are required to identify records that are similar and likely refer
to the same entity. Such quasi-identifiers (QID) are however often not
allowed to be shared between organisations due to privacy and
confidentiality concerns. 

Research in the area of \emph{privacy-preserving record linkage}
(PPRL) aims to develop techniques that facilitate the linking of
databases 
without the need for any sensitive data to be shared between the
organisations involved in the linkage process~\cite{Chr2020,Vat17}.
PPRL is conducted by
encoding or encrypting sensitive data of the database owners (DOs)
before being exchanged with other third party organisations 
(such as a linkage unit (LU)~\cite{Chr2020,Vat17}) to calculate the 
similarities between records. At the end of such a PPRL process, 
only limited information about those compared record pairs that 
were classified as matches is revealed to the DOs~\cite{Vat17}. 
Any PPRL technique must guarantee that no participating party can
learn anything about the sensitive data in any of the databases.
The PPRL process must also be secure such that no external
adversary can learn any sensitive information about the entities
in the databases that are being linked~\cite{Vat17}.

Any encoding or encryption method used in PPRL must facilitate
approximate similarity calculations between sensitive values
without the need for sharing the actual values~\cite{Vat17}.
Various techniques to securely calculate similarities between
values have been proposed. They either rely on expensive security
computations to achieve strong privacy guarantees, or they use
efficient data masking or perturbation techniques that, however,
can be vulnerable to cryptanalysis attacks~\cite{Chr17} that can re-identify
sensitive values in an encoded database.

In this paper, we propose a PPRL approach based on
a word embedding and binarisation~\cite{Tis19} to provide high linkage
quality and a high degree of privacy, where the approach consumes less
time for the comparison process. To the best of our knowledge, we are the first that use the embedding to generate a binary string of a record in the PPRL context.
We first generate a word embedding of each q-gram, 
resulting in a matrix of decimal numbers. The generated matrix is then
converted to a matrix of binary strings. For each record in a database,
we extract q-grams and their corresponding matrix of binary strings to 
generate the final binary string. The semi-trusted third party then 
compares the final binary string to find matches between databases. We 
evaluate our approach in terms of linkage quality, privacy, and time
complexity by using real-world data sets that contain only letters 
and mixed letters and digits.
\vspace{-0.5mm}
\section{Related Works}
\label{related_works}

Word embedding techniques have been used to find semantic similarity in natural language processing (NLP)~\cite{Cho21, Xia23}. Ye et al.~\cite{Ye16} proposed an approach for semantic matching between plaintext and code in the software engineering domain. They used the skip-gram model to predict tokens of code and words of plaintext. Kenter and Rijke~\cite{Ken15} conduct semantic matching between short plaintexts using the meta-features (differences of characteristics between word vectors) that are derived from the compared word vectors and the averaged word vectors of their word embeddings. However, it is a lack of order of words due to the use of meta-features, thus, it is not possible to be used when the order of words is required.

Susan et al.~\cite{Sus24} classify a match and unmatch between a resume and job profiles. They extract unique words based on the Maximum Entropy Partitioning (MEP) algorithm~\cite{Sus19} and transform these words into feature vectors using GloVe~\cite{Pen14} and Word2Vec~\cite{Chu17} word embeddings. These vectors are then passed to a Bidirectional long short-term memory (BiLSTM)~\cite{Ber22} to classify the documents. However, the results show that the highest accuracy was less than 80\%.

Tissier et al.~\cite{Tis19} proposed an approach to reduce memory usage in word embeddings of record values. They first create a word embedding matrix of a record value and then use the autoencoder to generate a binary vector. They calculate the similarity of a pair of record values and a pair of their corresponding binary vectors using the cosine similarity calculation~\cite{Chr12} and the Sokal and Michener similarity function~\cite{Sok58}, respectively. Their results show that the generated binary vectors provide the same performance for semantic similarity and classification tasks as conducted on the original real-values. However, the privacy of words was not in their consideration.

Abdalla et al.~\cite{Abd20} used word embedding to preserve the privacy of health information in clinical notes. They first use the continuous bag-of-words (CBOW)~\cite{Mik13} to generate word embeddings of clinical notes. They then replace tokens in the clinical note with the closest neighbouring tokens in the same embedding space.

Various techniques have been proposed in the PPRL context to preserve the sensitive information. The widely used technique is Bloom filter (BF) encoding~\cite{Sch09} because it is a set-based approach that allows approximate similarity calculation to classify a pair of records as matched or unmatched. However, Christen et al.~\cite{Chr17} have shown that the BF is vulnerable to a cryptanalysis attack.

Tabulation hashing (TabHash) and two-steps hashing (2SH) are other set-based encoding approaches. These approaches were proposed to provide a higher degree of privacy than the BF. The TabHash was proposed by Smith~\cite{Smi17}. The sets of tables are first generated, where each table in a set contains random bits. Each record value is encoded based on the defined number of hash functions, resulting in a hash value that will then be split and used as a key to select a random bit in the corresponding table. The selected bits are concatenated and used for the Jaccard similarity calculation~\cite{Chr12} to compare bits of records in a pair. The 2SH was proposed by Ranbaduge et al.~\cite{Ran20a}. The record value is encoded into a bit array (the first step), and the bits in the array are then encoded into integer numbers (the second step). The results of 2SH show that it provides a higher level of accuracy than the BF and TabHash. However, both TabHash and 2SH are vulnerable to the privacy attack proposed by Vidanage et al.~\cite{Vid20}.  

Yao et al.~\cite{Yao23} proposed a PPRL approach to create a combined Bloom Filters (CBF) of multiple attributes into a single BF, where the combination depends upon the type of record value which is whether a letter or digit. They also proposed a classification approach based on the Siamese Neural Network (SNN)~\cite{Nai10}. In their SNN-PPRL approach, the CBF of a record is first converted to a vector embedding whose features are then extracted using the BiLSTM~\cite{Ber22} and classified as a match and an unmatch.

Vatsalan et al.~\cite{Vat21} proposed an approach for data encoding and matching that data using counting BF and differential privacy. They first use a bag-of-words (BOW)~\cite{Gol17} to extract and count the frequency of words. They increment the counting BF by the number of frequencies, where the counting BF is then perturbed and added noise to guarantee differential privacy. Their approach provides high linkage quality and better utility compared to the bit vector based method~\cite{Zha19}.

Some discussed approaches used word embedding to match between record values. However, most of them do not concern the degree of privacy of sensitive data. Some discussed approaches proposed in the PPRL context that privacy is of concern, but they have been shown to be vulnerable to privacy attacks~\cite{Chr17, Vid20}. In our approach, we use the continuous bag-of-words (CBOW) and binarisation with the aim of providing high linkage quality and degree of privacy while using less record comparison time.


\section{Background}
\label{background}
In this section, we first describe the concept of word embedding. We then describe the binary encoding that is widely used in the PPRL context.


\subsection{Word Embedding}
\label{word_embedding}

The continuous bag-of-words (CBOW) was proposed by Mikolov et al.~\cite{Mik13}. It is one of the word embedding techniques that are often used for text classification and sentiment analysis in the natural language processing (NLP) context~\cite{Men20}. The CBOW is a feedforward neural network that contains an input layer, a projection layer, and an output layer~\cite{Liu20, Mik13}. The surrounding words of a target word, also known as history and future words, are used as an input layer~\cite{Mik13}.

The words in the input layer are first converted to one-hot encoded vectors, where the length of each vector equals the number of the vocabulary in the dataset~\cite{Cho21, Liu20}. The one-hot vectors are then transformed to a matrix of decimal numbers (word embeddings)~\cite{Cho21}, and the numbers in the matrix are used for the cumulative sum~\cite{Liu20} in the projection layer. The sum result is then used in the output layer to predict the target word~\cite{Cho21, Liu20, Mik13}.

Word2Vec is often referred to when working with word embeddings~\cite{Ma15}. It contains CBOW and Skip-gram models. In contrast to CBOW, the skip-gram predicts the surrounding words of a target word~\cite{Cho21}. While skip-gram has better learning than the CBOW, the CBOW outperforms the skip-gram in terms of computing speed~\cite{Ma15}. Therefore, in our approach, we use the CBOW to create word embeddings because the time used for creating word embeddings affects the overall time complexity of encoding of our approach.

\subsection{Binary Encoding}
\label{binary_encoding}

Binary vectors are commonly used in the PPRL applications due to
the ease of implementation and effectiveness in matching records.
One popular binary encoding technique that is both efficient and allows approximate matching is Bloom filter encoding. First applications
of Bloom filter (BF) based PPRL~\cite{Sch09} are now being employed in real-world
linkage applications across the world.
Recent research has however shown
that Bloom filters as used for PPRL can be vulnerable to
cryptanalysis attacks~\cite{Chr17} that can re-identify values encoded into sets
of BFs.

\begin{figure}[!t]
  \centering
  \includegraphics[width=0.5\textwidth]{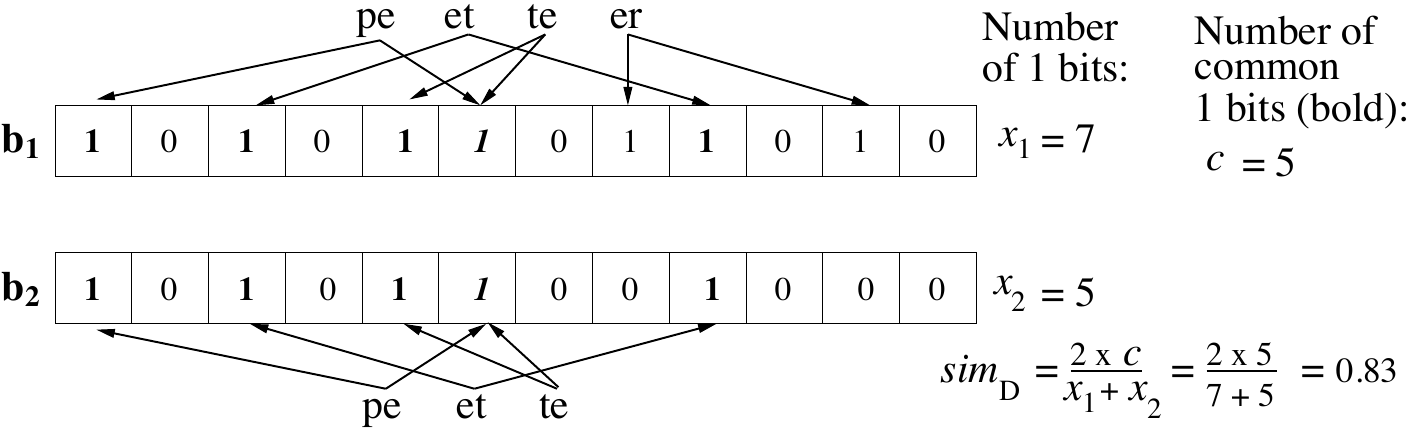}
  \caption{The Dice coefficient similarity~\cite{Chr12} calculation between the
    names \emph{``peter"} and \emph{``pete"}, converted into bigrams ($q=2$) and
    encoded into two Bloom filters $\mathbf{b}_1$ and $\mathbf{b}_2$
    of length $l=12$ bits using $k=2$ hash functions. The $1$ bits shown
    in italics at position $6$ is a hash collision, because both 
    \emph{``pe"} and \emph{``te"} are hashed to this position.}
    \label{fig:bloomfilter}
\end{figure}

Bloom filter (BF) encoding was proposed by Schnell et al.~\cite{Sch09} for PPRL 
because BFs can be used to efficiently calculate approximate similarities between records. A BF~\cite{Blo70}  $b$ is a bit vector of length $l= |b|$ where
initially all bits are set to 0. Each data information element in a set 
$s\in S$ is transformed into $l$ bits using $k> 1$ hash functions, 
where each hash function outputs an index value between 0 and $l- 1$. These
index values are then used to toggle the corresponding bits in vector $b$ to 1. 

In PPRL, the set $s$ is generally generated as q-grams, 
i.e., substrings of consecutive characters with a length $q$,  
from one or more quasi-identifying (QID) values from each record in a database, 
as shown in Fig.~\ref{fig:bloomfilter}. While various methods
have been proposed to encode strings~\cite{Dur12,Sch09,Vat13} as well as
numerical values using the BF encoding~\cite{Vat14c}, it has however been shown that the BF encoding
can be vulnerable to privacy attacks~\cite{Chr17, Chr18}. 
Sensitive values that occur 
frequently in an encoded database can lead to frequent bit patterns in BFs 
that can be identified~\cite{Chr2020}, and even individual frequent q-grams 
can be found using pattern mining techniques~\cite{Chr18}.



\begin{table*}[t!]
	\centering
	\caption{Common notation used in our approaches. \label{tab:notation}}
    \begin{footnotesize}
	\begin{tabular}{llll}
		\hline
		\noalign{\smallskip}
		$\mathbf{D}$ & \hspace{-3mm} Database & $\mathbf{Q}$ & \hspace{-3mm} Q-grams of a database \\
        $\mathbf{P}$ & \hspace{-3mm} A list of all possible q-grams &
        $\mathbf{E}$ & \hspace{-3mm} Encoded database \\
        $\mathbf{V}$ & \hspace{-3mm} A list of vectors of embeddings for all q-grams in $\mathbf{P}$ & $\mathbf{M}$ & \hspace{-3mm} A matrix of binaries of q-grams in $\mathbf{P}$ \\
        $\mathbf{M}'$ & \hspace{-3mm} A transpose matrix of $\mathbf{M}$ & $\boldsymbol{\phi}$ & \hspace{-3mm} Vectors of random binaries \\
        $\mathbf{T},t$ & \hspace{-3mm} A temporary binary string of $\mathbf{P}$ and each temporary binary string & $\mathbf{B}$ & \hspace{-3mm} Block of encoded database \\
        $\mathbf{B}_C$ & \hspace{-3mm} A common blocks & $v, vid$ & \hspace{-3mm} A record value and record identifier in a database \\ 
        $l_c$ & \hspace{-3mm} A length of the list of characters & $cbow$ & \hspace{-3mm} A CBOW model \\
        $q, \mathbf{q}$ & \hspace{-3mm} A length of q-gram and a list of q-grams & $c$ & \hspace{-3mm} A type of characters \\
        $d$ & \hspace{-3mm} A dimension of embeddings & $f_q$ & \hspace{-3mm} A minimum frequency of q-grams \\
        $w$ & \hspace{-3mm} A minimum number of q-grams before and after target q-gram; window size & $p_q$ & \hspace{-3mm} A possible q-gram in $\mathbf{P}$ \\
        $ep$ & \hspace{-3mm} A number of iterations & $s$ & \hspace{-3mm} A batch size \\ 
        $emb, emb_s$ & \hspace{-3mm} An embedding of $p_q$ and embedding with the batch size $s$ & $l$ & \hspace{-3mm} A length of binary string \\
        $reg_l$ & \hspace{-3mm} A regularisation loss & $rec_l$ & \hspace{-3mm} A reconstruction loss \\
        $k$ & \hspace{-3mm} A number of random bits & $l_f$ & \hspace{-3mm} A length of final binary string \\
        $\mathbf{q}_b$ & \hspace{-3mm} A list of binary string of a record &
        $q_v$ & \hspace{-3mm} A q-gram of a record value \\
        $b$ & \hspace{-3mm} A binary string of a record & $bkv$ & \hspace{-3mm} A blocking key value \\
        $sim$ & \hspace{-3mm} A similarity value & $sim_D$ & \hspace{-3mm} A Dice similarity \\
        $s_t$ & \hspace{-3mm} A similarity threshold & $|...|$ & \hspace{-3mm} Size or number of values in a database or list \\
        \noalign{\smallskip}
		\hline
		\noalign{\smallskip}
	\end{tabular}
    \end{footnotesize}
\end{table*}

\begin{figure}[t!]
  \centering
  \includegraphics[width=0.48\textwidth]{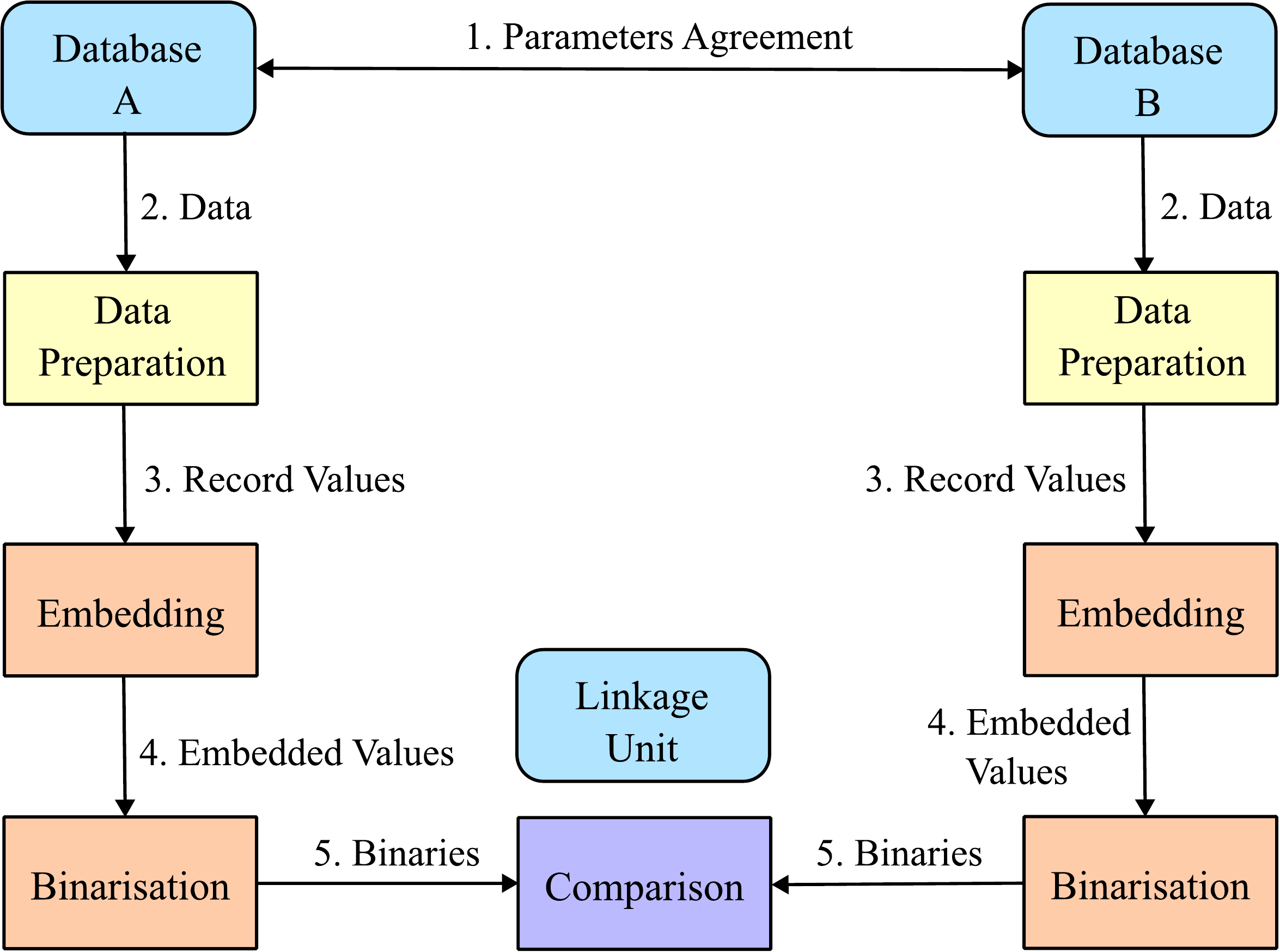}
  \caption{Overview protocol of our approach. The rounded blue boxes are the
  DOs' databases and the LU. The data preparation step is shown in yellow colour, while the encoding steps are shown in orange colour. The binary strings (encoded values) are sent to the LU for the comparison step which is shown in purple colour under the LU.}
  \label{fig:protocol_overview}
\end{figure}


\section{Methodology}
\label{methodology}

The notation used in our approach is as listed in Table~\ref{tab:notation}.
In our approach, we involve three parties, which are the two Database Owners (DOs) and a Linkage Unit (LU). We assume these parties follow the honest-but-curious (HBC) adversaries model~\cite{Hal10, Lin09}. The two DOs desire to share their records without any communication, except to make an agreement on parameter settings. The two DOs also desire to protect their sensitive information. Thus, the two DOs send their encoded record values to the LU to compare and find a match between records.

As illustrated in Fig.~\ref{fig:protocol_overview}, first before starting the data preparation step, the two DOs make an agreement on the parameters to be used in our protocol.
%
Each DO then prepares record values in its database by generating a list of q-grams of each record value. The DO also generates a list of all possible q-grams based on the type of characters. In the embedding and binarisation steps, the DO generates an embedding model (we use the CBOW model) and uses the generated model to embed each q-gram in the list of all possible q-grams. The DO then creates a random matrix and a random vector which are used with an embedding to generate binary strings of a q-gram. The DO maps binary strings with q-grams of each record in its database to generate a final binary string. The DO then generates blocks of encoded database before sending it to the LU. The LU conducts a comparison between the databases of the two DOs.


\subsection{Data Preparation Step}
\label{data_preparation}

\begin{figure}[t!]
    \centering
    \label{alg:alg1}
	 \begin{footnotesize}
	\setlength{\tabcolsep}{2pt}
	\begin{tabular}{lllll} \hline \noalign{\smallskip}
	\multicolumn{5}{l}{\textbf{Algorithm 1: \emph{Data preparation by a DO}}} \\ 
	\noalign{\smallskip} \hline \noalign{\smallskip}
	\multicolumn{5}{l}{Input:} \\
    - $\mathbf{D}$: & Database & - $q$: & Length of q-gram \\
    - $c$: & Type of characters \\
    \multicolumn{5}{l}{Output:} \\
    - $\mathbf{P}$: & List of all possible q-grams & - $\mathbf{Q}$: & Q-grams of a database \\
    1: & $\mathbf{P} \leftarrow [~]$ & 
    \multicolumn{3}{r}{// Initialise a list} \\
    2: & $\mathbf{Q} \leftarrow \{\}$ & 
    \multicolumn{3}{r}{// Initialise an inverted index} \\
    3: & $\mathbf{P} \leftarrow genAllPossQgrams(c,q)$ & \multicolumn{3}{r}{// Generate all possible q-grams} \\
 	4: & \textbf{for} $vid, v \in \mathbf{D}$ \textbf{do}: & 
    \multicolumn{3}{r}{// Loop over records in $\mathbf{D}$} \\
 	5: & ~~ $\mathbf{q} \leftarrow genQgramList(v,q)$ & 
    \multicolumn{3}{r}{// Generate a q-gram list of $v$} \\
    6: & ~~ $\mathbf{Q}[vid] \leftarrow \mathbf{q}$ & \multicolumn{3}{r}{// Insert a list $\mathbf{q}$ into $\mathbf{Q}$} \\
 	7: & return $\mathbf{P}, \mathbf{Q}$ & \\
	\noalign{\smallskip} \hline
    \end{tabular}
    \end{footnotesize}
\end{figure}

\begin{figure}[t!]
    \centering
	 \begin{footnotesize}
	\setlength{\tabcolsep}{2pt}
	\begin{tabular}{lllll} \hline \noalign{\smallskip}
	\multicolumn{5}{l}{\textbf{Algorithm 2: \emph{Encoding by a DO}}} \\ 
	\noalign{\smallskip} \hline \noalign{\smallskip}
	\multicolumn{5}{l}{Input:} \\
    \multicolumn{2}{l}{- $\mathbf{P}$:\hspace{0.5mm} List of all possible q-grams} & - $w$: & Size of window \\
    \multicolumn{2}{l}{- $f_q$: Minimum frequency of q-grams} & - $d$: & Embedding dimension \\
    \multicolumn{2}{l}{- $l$: \hspace{1mm} Length of binary string} & - $rec_l$: & Reconstruction loss \\
    \multicolumn{2}{l}{- $s$: \hspace{0.5mm} Batch size} & - $reg_l$: & Regularisation loss \\
    \multicolumn{2}{l}{- $ep$: Number of iterations} & - $k$: & Number of random bits \\
    \multicolumn{2}{l}{- $l_f$:\hspace{0.35mm} Length of final bit string} \\
    \multicolumn{5}{l}{Output:} \\
    \multicolumn{5}{l}{- $\mathbf{T}$: Inverted index of temporary binary strings} \\
    1: & \multicolumn{4}{l}{$\mathbf{V} \leftarrow \{\}$ \hspace{3.45cm} // Initialise an inverted index $\mathbf{V}$} \\
    2: & $\mathbf{M} \leftarrow \{\}\{\}$ & \multicolumn{3}{r}{// Initialise a matrix $\mathbf{M}$} \\
    3: & $\mathbf{M}_{emb} \leftarrow \{\}\{\}$ & \multicolumn{3}{r}{// Initialise a matrix $\mathbf{M}_{emb}$} \\
    4: & $\mathbf{M'} \leftarrow \{\}\{\}$ & \multicolumn{3}{r}{// Initialise a transpose of $\mathbf{M}$} \\
    5: & $\mathbf{M}_{P} \leftarrow \{\}\{\}$ & \multicolumn{3}{r}{// Initialise a matrix $\mathbf{M}_{P}$} \\
    6: & $\boldsymbol{\phi} \leftarrow < >$ & \multicolumn{3}{r}{// Initialise a vector $\boldsymbol{\phi}$} \\
    7: & \multicolumn{4}{l}{$\mathbf{T} \leftarrow \{\}$ \hspace{3.45cm} // Initialise an inverted index $\mathbf{T}$} \\
    8: & $cbow \leftarrow genModel(\mathbf{P}, d, f_q, w)$ & \multicolumn{3}{r}{// Generate CBOW model} \\
    9: & \textbf{for} $p_q \in \mathbf{P}$ \textbf{do}: & 
    \multicolumn{3}{r}{// Loop over q-gram in $\mathbf{P}$} \\
    10: & ~~ $emb \leftarrow getEmbCBOW(p_q)$ & \multicolumn{3}{r}{// Get embedding from CBOW} \\
    11: & ~~ $\mathbf{V}.add(emb)$ & \multicolumn{3}{r}{// Add embedding into $\mathbf{V}$} \\
    12: & $\mathbf{M} \leftarrow randMatrix(l, d)$ & \multicolumn{3}{r}{// Generate random $\mathbf{M}$}\\
    13: & $\boldsymbol{\phi} \leftarrow randVector(d)$ & \multicolumn{3}{r}{// Generate random $\boldsymbol{\phi}$} \\
    14: & \textbf{for} $i \ to \ ep$ \textbf{do}: & 
    \multicolumn{3}{r}{// Loop over $ep$} \\
    15: & ~~ \textbf{for} $j \ to \ |\mathbf{P}| - s \ step \ s$ \textbf{do}: & 
    \multicolumn{3}{r}{// Loop over interval of $s$} \\
    16: & \multicolumn{4}{l}{~~ ~~ $\mathbf{M} \leftarrow regGrad(\mathbf{M}, d, reg_l)$ \hspace{0.28cm}// Generate regularisation gradient} \\ 
    17: & ~~ ~~ $emb_s \leftarrow getEmbBat(\mathbf{V}, j, s)$ & \multicolumn{3}{r}{// Get embeddings with size $s$} \\
    18: & \multicolumn{4}{l}{\hspace{0.5cm} $\mathbf{M}, \boldsymbol{\phi} \leftarrow recGrad(\mathbf{M}, \boldsymbol{\phi}, emb_s, d, s, rec_l)$} \\
     \multicolumn{5}{r}{// Generate reconstruction gradient} \\
    19: & ~~ ~~ $reg_l, rec_l \leftarrow update(reg_l, rec_l)$ & \multicolumn{3}{r}{// Update $reg_l$ and $rec_l$} \\
    20: & \multicolumn{4}{l}{$\mathbf{M}_{emb} \leftarrow genEmbMat(\mathbf{V}, s, l)$ \hspace{0.56cm}// Generate matrix of embedding} \\
    21: & $\mathbf{M'} \leftarrow genTrans(\mathbf{M}) $ & \multicolumn{3}{r}{// Generate transpose of $\mathbf{M}$} \\
    22: & \multicolumn{4}{l}{$\mathbf{M}_\mathbf{P} \leftarrow genBinary(\mathbf{M}_{emb}, \mathbf{M'})$} \\
    \multicolumn{5}{r}{// Generate temporary binary string of $\mathbf{P}$} \\
 	23: & \textbf{for} $p_q \in \mathbf{P}$ \textbf{do}: & 
    \multicolumn{3}{r}{// Loop over $|\mathbf{P}|$} \\
    24: & ~~ $t \leftarrow genBin(\mathbf{P}, \mathbf{M}_\mathbf{P}, p_q, l, k, l_f)$  & \multicolumn{3}{r}{// Generate binary string of $p_q$} \\
    25: & ~~ $\mathbf{T}[\mathbf{P}[p_q]] \leftarrow t$ & \multicolumn{3}{r}{// Add $t$ into $\mathbf{T}[\mathbf{P}[p_q]]$} \\
    26: & return $\mathbf{T}$ \\
	\noalign{\smallskip} \hline
    \end{tabular}
    \end{footnotesize}
\end{figure}

In this step, each DO generates a list of all possible q-grams, $\mathbf{P}$, and an inverted index of q-grams $\mathbf{Q}$ for record values in a database, $\mathbf{D}$. Algorithm~1 outlines the data preparation step, where in lines 1 and 2, a DO first initialises a list of all possible q-grams $\mathbf{P}$ and an inverted index $\mathbf{Q}$. The inverted index $\mathbf{Q}$ is used for storing a list of q-grams of each record value in a database. In line 3, the DO uses the function $genAllPossQgrams()$ to generate a list of all possible q-grams based on the length of a q-gram, $q$, and the length of the list of characters, $l_c$, where a type of character, $c$, can be letters, digits, or a combination of letters and digits. The length of a list of all possible q-grams can be calculated as $|\mathbf{P}| = (l_c) ^ q$. For example, assuming that the length of a q-gram $q = 2$ and the two databases $\mathbf{D_A}$ and $\mathbf{D_B}$ of the two DOs, $DO_A$ and $DO_B$, respectively, contain only letters. Therefore, a list of all possible q-grams will be $\mathbf{P} = [aa, ab, ac, ..., zz]$ with the length $|\mathbf{P}| = 26^2 = 676$. 

In line 4, the DO loops over each record identifier $vid$ and record value $v$ in its database $\mathbf{D}$. The DO then uses the function $genQgramList()$ in line 5 to generate a list of q-grams of the record value $v$ where each q-gram is of the length $q$ resulting in a list of q-grams $\mathbf{q}$. In line 6, the DO adds the list $\mathbf{q}$ that corresponds to the record identifier $vid$ of $v$ into the inverted index $\mathbf{Q}[vid]$. The DO repeats the steps in lines 4 to 6 until the list of q-grams $\mathbf{q}$ of the last record in $\mathbf{D}$ is generated. This data preparation step provides the list of all possible q-grams $\mathbf{P}$ and an inverted index of a database $\mathbf{Q}$.


\begin{figure*}[t!]
  \centering
  \includegraphics[width=1.0\textwidth]{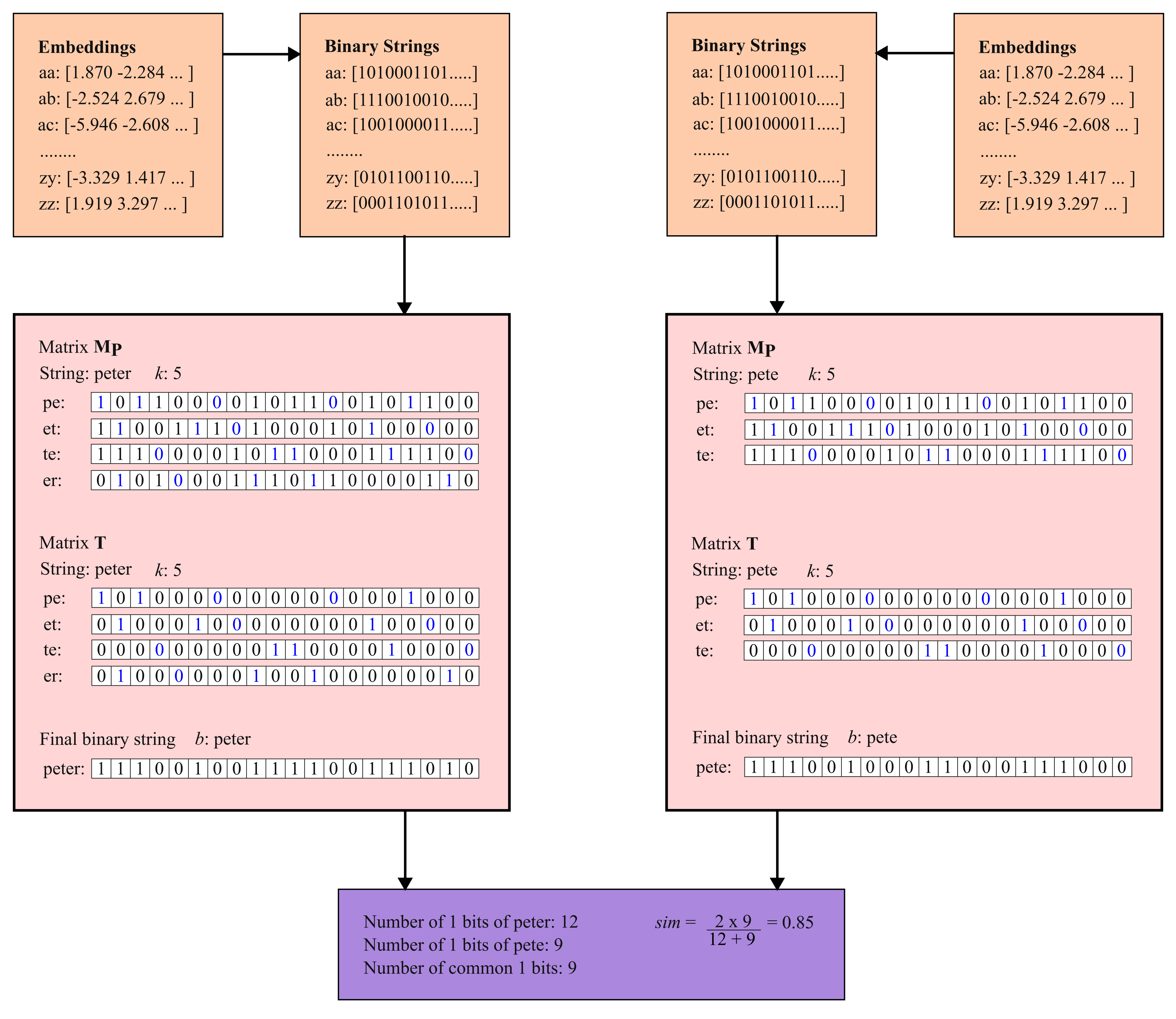}
  \caption{Example of encodings and comparison processes. The embeddings of all possible q-grams were first generated, and then each q-gram was encoded into a binary string. The embeddings and binary strings of all possible q-grams are shown in the orange boxes. The pink boxes show the binary string $peter$ of the first database and the binary string $pete$ of the second database. Each pink box shows the matrix $\mathbf{M}_\mathbf{P}$, the matrix $\mathbf{T}$, and the final binary string $b$ of the string, where each binary string was generated using $k = 5$. The final binary strings of the two databases are compared using the Dice similarity as shown in the purple box.}
  \label{fig:binary_example}
\end{figure*}


\subsection{Encoding Step}
\label{encoding}

In this step, a DO first uses the agreed dimension of embeddings $d$, the minimum frequency of q-grams $f_q$, the window size $w$, and the list of all possible q-grams, $\mathbf{P}$ to generate word embeddings. To generate word embeddings, the DO uses the Word2Vec based on the continuous bag-of-words (CBOW) model~\cite{Mik13}. However, any embedding technique can be used, and it can affect the linkage quality of the protocol.

As outlined in Algorithm~2, in lines 1 to 7, the DO initialises the inverted index $\mathbf{V}$ for storing the list of vectors of embeddings, the matrix $\mathbf{M}$ for storing random values with the size of $d \times l$ where the length of binary string $l > (l_c)^q$ ($l > |\mathbf{P}|$) to ensure the binary string can be generated from all embeddings in the matrix of size $|\mathbf{P}| \times l$, the matrix $\mathbf{M}_{emb}$ for storing all embeddings of $\mathbf{P}$, the transpose matrix $\mathbf{M'}$, the matrix $\mathbf{M}_\mathbf{P}$ for storing temporary binary strings of $\mathbf{P}$, the vector $\boldsymbol{\phi}$ for storing binary vectors where each with the length of $d$, and the inverted index $\mathbf{T}$ for storing temporary binary string to be used to generate the final binary string of each q-gram of a record. In line 8, the DO uses the list of all possible q-grams $\mathbf{P}$, the dimension of embeddings $d$, the minimum frequency of q-grams $f_q$, and the window size $w$ as inputs to the function $genModel()$ to generate the CBOW model, $cbow$.

In line 9, the DO loops over each possible q-gram $p_q$ in the list $\mathbf{P}$. The DO then uses the function $getEmbCBOW()$ to retrieve an embedding vector, $emb$, that corresponds to the possible q-gram $p_q$ in line 10. The DO then adds the $emb$ of $p_q$ into the list of vectors of embeddings $\mathbf{V}$ in line 11. The DO conducts the steps in lines 9 to 11 until there is no further q-gram $p_q$ in $\mathbf{P}$. Fig.~\ref{fig:binary_example} shows the examples of embeddings of all possible q-grams $p_q \in \mathbf{P}$ of the two DOs in the orange boxes.


Once the list $\mathbf{V}$ has been generated, the DO generates random binaries by using the function $randMatrix()$ and stores them in the matrix $\mathbf{M}$ in line 12. The DO then uses the function $randVector()$ to generate random binaries which are then stored in the vectors $\boldsymbol{\phi}$ in line 13.
In lines 14 and 15, the DO loops $ep$ iterations where in each iteration the DO loops the batch size $s$ in the range of the length of $\mathbf{P}$. The DO then uses the function $regGrad()$ to generate a regularisation gradient for the matrix $\mathbf{M}$ in line 16. After that, in line 17 the DO retrieves the embeddings with the size of $s$ by using the function $getEmbBat()$, and in line 18 the DO then uses the function $recGrad()$ to generate the reconstruction gradient for the matrix $\mathbf{M}$ and the vector $\boldsymbol{\phi}$. In line 19, the DO updates the regularisation and reconstruction losses using the function $update()$. The DO repeats the steps in lines 14 to 19 until the number of iterations $ep$ is reached. 

In line 20, the DO generates the matrix of embeddings, $\mathbf{M}_{emb}$ by using the function $genEmbMat()$. In line 21, the DO conducts the transpose of the matrix $\mathbf{M}$ by using the function $genTrans()$ resulting in the matrix $\mathbf{M'}$. The DO then uses the generated $\mathbf{M}_{emb}$ and $\mathbf{M'}$ as inputs into the function $genBinary()$ to generate a matrix of temporary binary strings of all q-grams in $\mathbf{P}$ resulting in the matrix $\mathbf{M}_\mathbf{P}$ in line 22. In the function $genBinary()$, the DO conducts the dot product between the $\mathbf{M}_{emb}$ and $\mathbf{M'}$. The DO then set the bit to 1 if the value in the $\mathbf{M}_\mathbf{P}$ is greater than 0, otherwise, the DO set the bit to 0. The examples of binary strings of q-grams in $p_q \in \mathbf{P}$ of the two DOs show in the orange boxes in Fig~\ref{fig:binary_example}.

\begin{figure}[t!]
    \centering
	 \begin{footnotesize}
	\setlength{\tabcolsep}{2pt}
	\begin{tabular}{lllll} \hline \noalign{\smallskip}
	\multicolumn{5}{l}{\textbf{Algorithm 3: \emph{Generating final binary string by a DO}}} \\ 
	\noalign{\smallskip} \hline \noalign{\smallskip}
	\multicolumn{5}{l}{Input:} \\
    - $\mathbf{Q}$: & Q-grams of a database & - $l_f$: & Length of final binary string \\
    - $\mathbf{T}$: & Temporary binary strings &  \\
    \multicolumn{5}{l}{Output:} \\
    - $\mathbf{B}$: & Blocks of an encoded database \\
    1: & $\mathbf{B} \leftarrow \{\}$ & 
    \multicolumn{3}{r}{// Initialise an inverted index of blocks} \\
    2: & $\mathbf{E} \leftarrow \{\}$ & 
    \multicolumn{3}{r}{// Initialise an encoded database} \\
    3: & \textbf{for} $vid, \mathbf{q} \in \mathbf{Q}$ \textbf{do}: & 
    \multicolumn{3}{r}{// Loop over q-gram list in $\mathbf{Q}$} \\
    4: & ~~ $\mathbf{q}_b = [ \ ]$ & 
    \multicolumn{3}{r}{// Initialise binary string of record} \\
    5: & ~~ \textbf{for} $q_v \in \mathbf{q}$ \textbf{do}: & 
    \multicolumn{3}{r}{// Loop over q-gram in $\mathbf{q}$} \\
    6: & ~~ ~~ $t \leftarrow \mathbf{T}[q_v]$ & \multicolumn{3}{r}{// Get binary of a q-gram} \\
 	7: & ~~ ~~ $\mathbf{q}_b.add(t)$ & 
    \multicolumn{3}{r}{// Add q-gram binary to $\mathbf{q}_b$} \\
 	8: & ~~ $b \leftarrow genFinalBin(\mathbf{q}_b, l_f)$ & 
    \multicolumn{3}{r}{// Generate final binary string of record} \\
    
    9: & ~~ $\mathbf{E}[vid] \leftarrow b$ & 
    \multicolumn{3}{r}{// Add the final binary string to $\mathbf{E}[vid]$} \\
 	10: & $\mathbf{B} \leftarrow genBlocks(\mathbf{E})$ & 
    \multicolumn{3}{r}{// Generate blocks of $\mathbf{E}$} \\
    11: & return $\mathbf{E}$ & \\
	\noalign{\smallskip} \hline
    \end{tabular}
    \end{footnotesize}
\end{figure}

In line 23, the DO loops over each q-gram $p_q \in \mathbf{P}$. The DO then uses the function $genBin()$ to generate a temporary binary string $t$ in line 24. In this $genBin()$ function, the DO uses $p_q$ as a random seed to randomly select $k$ bits to generate the temporary binary string $t$. For the unselected bits, the DO sets the bit to 0s. However, when the type of characters is the combination of letters and digits, the length $l$ can be longer than the final length of the binary string $l_f$, $l > l_f$. In this case, after the temporary binary string $t$ has been generated, the DO uses $\mathbf{P}[p_q]$ as a seed to select $k$ bits again from $t$. For example, assume that the final length of the binary string is $l_f = 1,000$, the length of q-grams $q = 2$, and the combination of letters and digits is $l_c = 36$. The $|\mathbf{P}| = 36^2 = 1,296$ which is longer than $l_f$. Therefore, the DO will need to randomly select bits to create the final length of the binary string as required which is 1,000.

For example, as illustrated in Fig.~\ref{fig:binary_example} in the pink boxes, assume the two strings of the $DO_A$ and $DO_B$ are \emph{``peter"} and \emph{``pete"}, respectively, and the defined $k = 5$. The randomly selected bits of each q-gram are shown in blue colour. If the selected bits in the matrix $\mathbf{M}_\mathbf{P}$ are 1s, the bits are set to 1s in the matrix $\mathbf{T}$, otherwise, the bits in the matrix $\mathbf{T}$ are set to 0s. Therefore, from the binary string \emph{``pe"} of the strings \emph{``peter"} and \emph{``pete"} in the matrix $\mathbf{M}_\mathbf{P}$ is \emph{``pe"} $= ``10110000101100101100"$ will be set to \emph{``pe"} $= ``10100000000000001000"$ in the matrix $\mathbf{T}$.

Once $t$ is created, the DO then inserts a q-gram $\mathbf{P}[i]$ as a key and inserts $t$ as a value into the inverted index $\mathbf{T}$ in line 26. The DO repeats the steps in lines 24 to 26 until the binary string of the last embedding in the $\mathbf{M}_\mathbf{P}$ is generated. The inverted index of temporary binary strings $\mathbf{T}$ is then returned as the output in line 27.

Algorithm~3 outlines the mapping of binary strings from $\mathbf{T}$ to each q-gram value of a record value in the database to create the final binary string of the record. In lines 1 and 2, the DO initialises an encoded database $\mathbf{E}$. The DO then loops over the inverted index of q-grams $\mathbf{Q}$ in line 3 to retrieve a record identifier and its corresponding list of q-grams. The DO initialises a list of q-gram binary strings $\mathbf{q}_b$ in line 4. In lines 5 and 6, the DO loops over the list of q-grams $\mathbf{q}$ to extract the temporary binary string, $t$, of each q-gram of the record from $\mathbf{T}$. After that, in line 7, the DO adds $t$ into the list $\mathbf{q}_b$.

Once all temporary binary strings that correspond to the record are extracted, the DO uses the function $genFinalBin()$, the list $\mathbf{q}_b$, and the final length $l_f$ to generate the final binary string $b$ of the record in line 8. In the function $genFinalBin()$, the DO creates a matrix of size $|\mathbf{q}| \times l_f$ where each row refers to each q-gram and each column refers to each bit in the binary string. For each column, if a 1-bit is found, the DO sets 1 to the corresponding bit position in $b$, otherwise, the DO sets 0 to the bit position in $b$. As illustrated in Fig.~\ref{fig:binary_example} in the bottom of the pink boxes, to generate the final binary string $b$, if the 1 bits in $\mathbf{T}$ are found, the bits in the corresponding positions in $b$ are set to 1. In line 9, the DO adds $b$ that corresponds to the record identifier $vid$ into the encoded database $\mathbf{E}$. In line 10, the DO uses the function $genBlocks()$ to generate blocks of binary strings, $\mathbf{B}$, of the encoded database $\mathbf{E}$. This is to reduce the comparison space when the Linkage Unit (LU), a semi-trusted third party, conducts the comparison process.




\subsection{Comparison Step}
\label{comparison}

\begin{figure}
\begin{center}
  \label{algo_lu}
 \begin{footnotesize}
  \setlength{\tabcolsep}{2pt}
  \begin{tabular}{ll} \hline \noalign{\smallskip}
  \multicolumn{2}{l}{\textbf{Algorithm 4: \emph{Linking
    Encoded Databases}}} \\ 
  \noalign{\smallskip} \hline \noalign{\smallskip}
  \multicolumn{2}{l}{Input:} \\
  \multicolumn{2}{l}{- $\mathbf{B}_A$:\hspace{2mm} Blocks of encoded database from the $DO_A$} \\
  \multicolumn{2}{l}{- $\mathbf{B}_B$:\hspace{2mm} Blocks of encoded database from the $DO_B$} \\
  \multicolumn{2}{l}{- $s_t$:\hspace{4mm} Similarity threshold} \\
  \multicolumn{2}{l}{Output:} \\
    \multicolumn{2}{l}{- $\mathbf{R}$:\hspace{3mm} Matched record
      pairs} \\
    \noalign{\smallskip}
  1:  & $\mathbf{R} = \{\}$ \hspace*{\fill} // Initialise inverted
        index of matches \\         
  2:  & $\mathbf{B}_c = \mathbf{B}_A \cap \mathbf{B}_B$
        \hspace*{\fill} // Get the common blocks \\ 
  3:  & \textbf{for} $bkv \in \mathbf{B}_c$ \textbf{do}:
        \hspace*{\fill} // Loop over common blocks \\
  4:  & ~~ \textbf{for} $(vid_A, b_A) \in
        \mathbf{B}_A[bkv]$ \textbf{do}: \hspace*{\fill} //
        Loop over $\mathbf{B}_A[bkv]$ \\
  5:  & ~~ ~~ \textbf{for} $(vid_B, b_B) \in
        \mathbf{B}_B[bkv]$ \textbf{do}: \hspace*{\fill} //
        Loop over $\mathbf{B}_B[bkv]$ \\
  6:  & ~~ ~~ ~~ $sim = sim_D(b_A, b_B)$ \hspace*{\fill} ~~ //
        Dice similarity calculation \\
  7:  & ~~ ~~ ~~ \textbf{if} $sim \geq s_t$ \textbf{do}:
        \hspace*{\fill} // Check if record pair is a match \\
  8: & ~~ ~~ ~~ ~~ $\mathbf{R}[(vid_A, vid_B)] = sim$
       \hspace*{\fill} // Add record pair to matches \\
  9: & return $\mathbf{R}$ \hspace*{\fill} // Send matched record
        pairs to DOs \\
  \hline
  \end{tabular}
  \end{footnotesize}
  \end{center}
\end{figure}

The LU receives blocks of encoded database, $\mathbf{B}_A$ and $\mathbf{B}_B$, and the similarity threshold, $s_t$, from the $DO_A$ and $DO_B$, respectively. As outlined in Algorithm~4, in line 1, the LU first initialises the inverted index $\mathbf{R}$ for storing matched encoded record pairs. The LU then finds common blocks, $\mathbf{B}_C$, between $\mathbf{B}_A$ and $\mathbf{B}_B$ in line 2. For each blocking key value, $bkv$, that corresponds to a common block in $\mathbf{B}_C$, in lines 4 and 5 the LU loops over $\mathbf{B}_A$ and $\mathbf{B}_B$ to extract the record identifiers $vid_A$ and $vid_B$, and binary strings $b_A$ and $b_B$, respectively. The LU then uses the function $sim_D()$ to calculate the Dice coefficient similarity~\cite{Chr12} between binary strings $b_A$ and $b_B$, resulting in the similarity value $sim$ in line 6. In line 7, the LU checks whether $sim$ is at least $s_t$. 

For example, as illustrated in Fig.~\ref{fig:binary_example} in the purple box, the Dice coefficient similarity~\cite{Chr12} calculation between the binary strings \emph{``peter"} $= ``11100100111100111010"$ and \emph{``pete"} $= ``11100100011000111000"$ is $sim = 0.85$. The number of 1 bits of \emph{``peter"} is 12 and the number of 1 bits of \emph{``pete"} is 9, while the number of common 1 bits is 9. The Dice coefficient similarity calculation is

\begin{footnotesize}
\begin{equation*}
    sim = \frac{2 \times {common \ 1 \ bits}}{summation \ of \ numbers \ of \ 1 \ bits \ of \ the \ two \ binary \ strings}.
\end{equation*}
\end{footnotesize}

Therefore, the $sim$ of the two binary strings $\emph{``peter"}$ and $\emph{``pete"}$ is $sim = (2 \times 9)/(12 + 9) = 18/21 = 0.85$. In line 8, if $sim$ of the pair of binary strings is $sim \geq s_t$, the LU adds the pair of record identifiers as a key and adds $sim$ as a value into the inverted index of matched records $\mathbf{R}$. The LU repeats the steps in lines 3 to 8 until no further blocks are to be compared. The LU then returns $\mathbf{R}$ to the two DOs.


\section{Theoretical Analysis}
\label{theoretical_analysis}
In this section, we provide a theoretical analysis of our approach in 
terms of complexity, linkage quality, and privacy. 

\subsection{Complexity Analysis}
\label{complexity}

As shown in Algorithm~1, each DO first generates a list
of all possible q-grams $\mathbf{P}$ where the length of $\mathbf{P}$, $|\mathbf{P}|$ depends upon a list of characters $l_c$ and a length of a q-gram $q$. Therefore, the DO requires $O((l_c)^q)$ time complexity to generate $\mathbf{P}$. The DO then generates an inverted index of q-grams $\mathbf{Q}$ of records of their database $\mathbf{D}$. Assuming each record $v\in\mathbf{D}$ contains $n$ q-grams, to generate $\mathbf{Q}$, the DO requires a complexity of $O(n \times |\mathbf{D}|)$.

To generate binary strings of a database, as shown in Algorithm~2, the DO first generates an embedding model $cbow$, which requires a time complexity of $O(|\mathbf{P}|)$. The DO extracts an embedding for each q-gram $p_q \in \mathbf{P}$. In this step, the DO also requires $O(|\mathbf{P}|)$ time complexity. The DO then generates a random matrix $\mathbf{M}$ and a random vector $\phi$. In these steps, the DO requires $O(l \times d)$ and $O(d)$ for generating $\mathbf{M}$ and $\phi$, respectively, where $l$ is the length of a binary string and $d$ is an embedding dimension. Once $\mathbf{M}$ and $\phi$ are generated, the DO loops $ep$ times over $\mathbf{P}$ to generate and update regularisation gradient and reconstruction gradient, where $ep$ is agreed by the two DOs. In this step, the DO requires $O(ep \times |\mathbf{P}|)$ time complexity. The DO then generate the matrix of embeddings $\mathbf{M}_{emb}$ which requires $O|\mathbf{V}|$ time complexity. After that, the DO creates a transpose of the matrix $\mathbf{M}$ requiring a time complexity of $O(|\mathbf{M}|)$. In the last step, to generate a temporary binary string of the database $\mathbf{T}$, the DO requires a time complexity of $O(|\mathbf{P}|)$.

In the mapping binaries to record values step shown in Algorithm~3, the DO first extracts a list of q-grams $\mathbf{q}$ of a record from the q-grams of the database $\mathbf{Q}$. For each q-gram in the list $\mathbf{q}$, the DO extracts its corresponding binary strings from the temporary binary strings $\mathbf{T}$. The DO concatenates binary strings of a record into a single binary string and stores it in the inverted index of the encoded database $\mathbf{E}$. Assuming each record contains $n$ q-grams, in these steps the DO requires $O(|\mathbf{Q}| \times (n \times |\mathbf{T}|))$ time complexity. Once the DO generates binary strings of every record in its database, the DO generates blocks of binary strings $\mathbf{B}$ which requires $O(\mathbf{E})$.

As shown in Algorithm~4, the comparison step by the LU, the LU requires $O(\mathbf{B}_A \times \mathbf{B}_B)$ time complexity to find the common blocks of encodings $\mathbf{B}_C$ between $\mathbf{B}_A$ and $\mathbf{B}_B$ receiving from the $DO_A$ and $DO_B$, respectively. The LU then loops over $\mathbf{B}_C$ to retrieve each common block. The LU extracts binary strings in each common block. We assume the number of binary strings in every block of the $DO_A$ equals $n$ and $DO_B$ equals $m$. Therefore, the LU requires the time complexity of $O(|\mathbf{B}_C| \times n^m)$ for comparing between binary strings of the two DOs.




\subsection{Linkage Quality Analysis}
\label{linkage_quality}


The linkage quality of our approach depends upon the embedding process where any embedding technique can be used, the random $k$ value, the length of a record value, and the algorithm for generating blocks. The embedding process affects the linkage quality because it is the first step of binarisation where a bit in a binary string of a q-gram will be set to either 1 or 0 depending upon the embedding values. As described in Section~\ref{encoding}, a bit is set to 1 if the value in the matrix $\mathbf{M}_\mathbf{P}$ is greater than 0, otherwise, the bit is set to 0.

The agreed random $k$ value affects the linkage quality of our approach because the larger $k$ can possibly result in more number of 1 bits, thus, the final binary string will contain many 1 bits leading to more false positives. Similarly, the longer length of the record value has a higher possibility that the final binary string will contain many 1 bits and leading to more false positives.
%
The algorithm for generating blocks also affects the linkage quality. This is because similar record values will not be compared if they are inserted into different blocks, thus, a higher number of false negatives. The linkage quality results are shown in Tables~\ref{tab:ncvr_linkquality} and~\ref{tab:acmscholar_linkquality}.


\subsection{Privacy Analysis}
\label{privacy}


We assume the LU is a semi-honest adversary that wants to learn the plaintext record values of the two DOs. The two DOs first agree on the parameter settings, thus, the DO learns about parameters that are used in both DOs. However, the DOs cannot learn any sensitive information about each other. The DOs then individually generate all possible q-grams using the agreed length of q-gram and the type of characters. This allows them to learn the list of all possible q-grams. The DOs cannot learn anything about q-grams of each database because they do not share any information.

In the encoding step, the DOs individually generate embeddings and binary strings of all possible q-grams. In this step, both DOs learn an embedding and a binary string of each q-gram, but they cannot learn any sensitive value of the other because they cannot know which q-gram is contained in another database. In the generating final binary string step, the DOs generate a final binary string of the q-grams corresponding to the record values in their databases. The DOs cannot learn anything about the other because they do not share any value related to their databases. The DOs then generate blocks of their binary strings. In this step, both DOs also cannot learn anything about each other because the blocks are generated based on the record value of each database.

The DOs send their encoded databases to a LU. The LU first finds common blocks between databases. The LU can learn the common and not common blocks of encodings of the two databases, but it cannot learn the plaintext value encoded in them. This is because the multiple steps in generating binary strings make it difficult to reidentify the original record values. However, if we assume the LU knows the steps of encoding and generating binary strings, it is still difficult for the LU to reidentify the original records. This is because in the generating final binary string step, the 1 bits of each binary string are set based on bits in the same position of multiple q-grams in a record. Therefore, without knowing the plaintext of a record value, the LU cannot learn any sensitive information. In the last step of the comparison process, the LU returns pairs of record identifiers and their corresponding similarity values to the DOs. The DOs cannot learn anything except whether its records are matched or unmatched.

\section{Experimental Study}
\label{experimental_study}

We evaluated our approach (EmbBin) compared to three baselines, which are Bloom Filter (BF) encoding~\cite{Sch09}, Tabulation Hash (TabHash)~\cite{Smi17}, and Two-steps Hashing (2SH)~\cite{Ran20a} in terms of linkage quality, time complexity, and degree of privacy. Our approach is compared to the BF because the BF is considered a standard PPRL technique and it is binary string based. Similar to the BF, we compared our approach to the TabHash because it is binary string based, while we compared our approach to the 2SH because it is binary based before being encoded into other hash values.

We implemented our approach and the three baselines using Python 3.12 and ran experiments on a server with a minimum 0.8 GHz and maximum 4.0 GHz CPUs running on Ubuntu 24.04.

\subsection{Datasets and Parameter Setup}
\label{datasets_setup}

We used real-world data from the North Carolina Voter Registration\footnote{\url{http://dl.ncsbe.gov/}} (NCVR)~\cite{Ran20, Vai22, Vid23} and the DBLP computer science bibliograph (DBLP). For the NCVR datasets, we extracted attributes first name (FN), first and last names (FN and LN), first, last, and street address names (FN, LN, and SA), and first, last, street address, and city names (FN, LN, SA, and CT). For the DBLP datasets, we extracted the title and venue of ACM, DBLP1, and DBLP2. We used 251,294 records for each of the NCVR datasets and used 2,294 for ACM and 2,616 records for DBLP1 and DBLP2. We used these datasets as the first dataset in a pair. We then corrupted each of these datasets by 20\% of the length of each record and used them as the second dataset in a pair. Overall, we evaluated our approach and the baselines on seven dataset pairs.

For the parameter settings, we set the type of characters $c = alphabet$ and the length of characters $l_c = 26$ for the FN, and FN and LN of NCVR datasets while we set $c = mix$ and $l_c = 36$ for the other two NCVR datasets and all of the DBLP datasets, where mix is a combination of letters (the length of characters is 26) and digits (the length of characters is 10). We generated a list of q-grams $\mathbf{q}$ using the length of q-gram $q = 2$ for all datasets. We set the dimension of embedding $d = 300$, the minimum frequency of q-grams $f_q = 1$, and the minimum number of q-grams before and after the target q-gram (window size) $w = 5$. To generate a binary string of each q-gram, for the length of the binary string $l$, we ensure the binary string can be generated from all embeddings in the matrix of size $(l_c) ^ q \times l$, and thus we set $l > (l_c) ^ q$. Therefore, we set $l = 1,000$ for $c = alphabet$ and set $l = 2,000$ for $c = mix$. We set the number of random bits $k = 15$, the batch size $s = 75$, and the number of iterations $ep = 5$. For the final length of binary string $l_f$,  we set $l_f = 1,000$. For the similarity threshold $s_t$, we used $s_t = [0.8, 0.9, 1.0]$ to classify the matched and unmatched binary string pairs.

\subsection{Linkage Quality Results}
\label{linkage_quality_results}

\begin{table*}[t!]
  \centering
  \caption{Precision, recall, accuracy, and F1 for different NCVR dataset pairs evaluated on different approaches. The worst results of each dataset pair on different approaches are shown in bold italic.
  \label{tab:ncvr_linkquality}}
  \setlength{\tabcolsep}{2pt}
  \begin{tabular}{ccccccccccccccccccccccccc}
  \hline\noalign{\smallskip}
  Similarity & & \multirow{2}{*}{Measures} & & \multicolumn{4}{c}{FN} & & \multicolumn{4}{c}{FN and LN} & & \multicolumn{4}{c}{FN, LN, and SA} & & \multicolumn{4}{c}{FN, LN, SA, and CT} \\
  Threshold & & & & EmbBin & BF & TabHash & 2SH & & EmbBin & BF & TabHash & 2SH & & EmbBin & BF & TabHash & 2SH & & EmbBin & BF & TabHash & 2SH \\
  \noalign{\smallskip}\hline
  \noalign{\smallskip}
  \multirow{4}{*}{$s_t = 0.8$} & & Precision & & 0.9 & 0.99 & \textit{\textbf{0.77}} & 1.0 & & 0.99 & 0.99 & \textit{\textbf{0.67}} & 1.0 & & 0.86 & 0.96 & \textit{\textbf{0.33}} & 1.0 & & \textbf{\textit{0.57}} & 0.95 & 0.83 & 1.0 \\
  & & Recall & & 0.9 & 0.97 & 1.0 & \textbf{\textit{0.86}} & & 0.96 & 1.0 & 1.0 & \textbf{\textit{0.92}} & & 0.98 & 1.0 & 1.0 & \textbf{\textit{0.93}} & & 1.0 & 1.0 & 1.0 & \textbf{\textit{0.95}} \\
  & & Accuracy & & 1.0 & 1.0 & 1.0 & 1.0 & & 1.0 & 1.0 & 1.0 & 1.0 & & 1.0 & 1.0 & 1.0 & 1.0 & & 1.0 & 1.0 & 1.0 & 1.0 \\
  & & F1 & & 0.9 & 0.98 & \textbf{\textit{0.87}} & 0.92 & & 0.97 & 0.99 & \textbf{\textit{0.8}} & 0.96 & & 0.92 & 0.98 & \textbf{\textit{0.5}} & 0.96 & & \textbf{\textit{0.73}} & 0.97 & 0.91 & 0.97 \\
  \noalign{\smallskip}
  \hdashline
  \noalign{\smallskip}
  \multirow{4}{*}{$s_t = 0.9$} & & Precision & & 0.92 & 0.98 & \textbf{\textit{0.86}} & 1.0 & & 1.0 & 1.0 & 1.0 & 1.0 & & \textbf{\textit{0.99}} & 1.0 & 1.0 & 1.0 & & \textbf{\textit{0.98}} & 1.0 & 1.0 & 1.0 \\
  & & Recall & & \textbf{\textit{0.96}} & 1.0 & 0.99 & \textbf{\textit{0.96}} & & 1.0 & 1.0 & 1.0 & 1.0 & & 1.0 & 1.0 & 1.0 & 1.0 & & 1.0 & 1.0 & 1.0 & 1.0 \\
  & & Accuracy & & 1.0 & 1.0 & 1.0 & 1.0 & & 1.0 & 1.0 & 1.0 & 1.0 & & 1.0 & 1.0 & 1.0 & 1.0 & & 1.0 & 1.0 & 1.0 & 1.0 \\
  & & F1 & & 0.94 & 0.99 & \textbf{\textit{0.92}} & 0.98 & & 1.0 & 1.0 & 1.0 & 1.0 & & \textbf{\textit{0.99}} & 1.0 & 1.0 & 1.0 & & 1.0 & 1.0 & 1.0 & 1.0 \\
  \noalign{\smallskip}
  \hdashline
  \noalign{\smallskip}
  \multirow{4}{*}{$s_t = 1.0$} & & Precision & & 1.0 & 1.0 & 1.0 & 1.0 & & 1.0 & 1.0 & 1.0 & 1.0 & & 1.0 & 1.0 & 1.0 & 1.0 & & 1.0 & 1.0 & 1.0 & 1.0 \\
  & & Recall & & 1.0 & 1.0 & 1.0 & 1.0 & & 1.0 & 1.0 & 1.0 & 1.0 & & 1.0 & 1.0 & 1.0 & 1.0 & & 1.0 & 1.0 & 1.0 & 1.0 \\
  & & Accuracy & & 1.0 & 1.0 & 1.0 & 1.0 & & 1.0 & 1.0 & 1.0 & 1.0 & & 1.0 & 1.0 & 1.0 & 1.0 & & 1.0 & 1.0 & 1.0 & 1.0 \\
  & & F1 & & 1.0 & 1.0 & 1.0 & 1.0 & & 1.0 & 1.0 & 1.0 & 1.0 & & 1.0 & 1.0 & 1.0 & 1.0 & & 1.0 & 1.0 & 1.0 & 1.0 \\
  \noalign{\smallskip}
  \hline
  \end{tabular}
\end{table*}

\begin{table*}[t!]
  \centering
  \caption{Precision, recall, accuracy, and F1 for ACM, DBLP1, and DBLP2 dataset pairs evaluated on different approaches. The worst results of each dataset pair on different approaches are shown in bold italic.
  \label{tab:acmscholar_linkquality}}
  \setlength{\tabcolsep}{2pt}
  \begin{tabular}{ccccccccccccccccccccc}
  \hline\noalign{\smallskip}
  Similarity & & \multirow{2}{*}{Measures} & & \multicolumn{4}{c}{ACM} & & \multicolumn{4}{c}{DBLP1} & & \multicolumn{4}{c}{DBLP2} \\
  Threshold & & & & EmbBin & BF & TabHash & 2SH & & EmbBin & BF & TabHash & 2SH & & EmbBin & BF & TabHash & 2SH \\
  \noalign{\smallskip}\hline
  \noalign{\smallskip}
  \multirow{4}{*}{$s_t = 0.8$} & & Precision & & \textbf{\textit{0.55}} & 0.64 & 0.73 & 1.0 & & 0.98 & 0.98 & \textbf{\textit{0.96}} & 1.0 & & 0.97 & 0.98 & \textbf{\textit{0.96}} & 1.0 \\
  & & Recall & & 1.0 & 1.0 & 1.0 & \textbf{\textit{0.91}} & & 1.0 & 1.0 & 1.0 & \textbf{\textit{0.97}} & & 1.0 & 1.0 & 1.0 & \textbf{\textit{0.97}} \\
  & & Accuracy & & 1.0 & 1.0 & 1.0 & 1.0 & & 1.0 & 1.0 & 1.0 & 1.0 & & 1.0 & 1.0 & 1.0 & 1.0 \\
  & & F1 & & \textbf{\textit{0.71}} & 0.78 & 0.84 & 0.95 & & 0.99 & 0.99 & \textbf{\textit{0.98}} & \textbf{\textit{0.98}} & & \textbf{\textit{0.98}} & 0.99 & \textbf{\textit{0.98}} & \textbf{\textit{0.98}} \\
  \noalign{\smallskip}
  \hdashline
  \noalign{\smallskip}
  \multirow{4}{*}{$s_t = 0.9$} & & Precision & & \textbf{\textit{0.98}} & \textbf{\textit{0.98}} & 0.99 & 1.0 & & \textbf{\textit{0.99}} & \textbf{\textit{0.99}} & 1.0 & 1.0 & & \textbf{\textit{0.99}} & \textbf{\textit{0.99}} & \textbf{\textit{0.99}} & 1.0 \\
  & & Recall & & 1.0 & 1.0 & 1.0 & \textbf{\textit{0.99}} & & 1.0 & 1.0 & 1.0 & \textbf{\textit{0.99}} & & 1.0 & 1.0 & 1.0 & \textbf{\textit{0.99}} \\
  & & Accuracy & & 1.0 & 1.0 & 1.0 & 1.0 & & 1.0 & 1.0 & 1.0 & 1.0 & & 1.0 & 1.0 & 1.0 & 1.0 \\
  & & F1 & & 0.99 & 0.99 & 0.99 & 0.99 & & \textbf{\textit{0.99}} & \textbf{\textit{0.99}} & 1.0 & \textbf{\textit{0.99}} & & 0.99 & 0.99 & 0.99 & 0.99 \\
  \noalign{\smallskip}
  \hdashline
  \noalign{\smallskip}
  \multirow{4}{*}{$s_t = 1.0$} & & Precision & & 1.0 & 1.0 & 1.0 & 1.0 & & 1.0 & 1.0 & 1.0 & 1.0 & & 1.0 & 1.0 & 1.0 & 1.0 \\
  & & Recall & & 1.0 & 1.0 & 1.0 & 1.0 & & 1.0 & 1.0 & 1.0 & 1.0 & & 1.0 & 1.0 & 1.0 & 1.0 \\
  & & Accuracy & & 1.0 & 1.0 & 1.0 & 1.0 & & 1.0 & 1.0 & 1.0 & 1.0 & & 1.0 & 1.0 & 1.0 & 1.0 \\
  & & F1 & & 1.0 & 1.0 & 1.0 & 1.0 & & 1.0 & 1.0 & 1.0 & 1.0 & & 1.0 & 1.0 & 1.0 & 1.0 \\
  \noalign{\smallskip}
  \hline
  \end{tabular}
\end{table*}

\begin{figure*}[t!]
  \centering
  \includegraphics[width=0.89\textwidth]{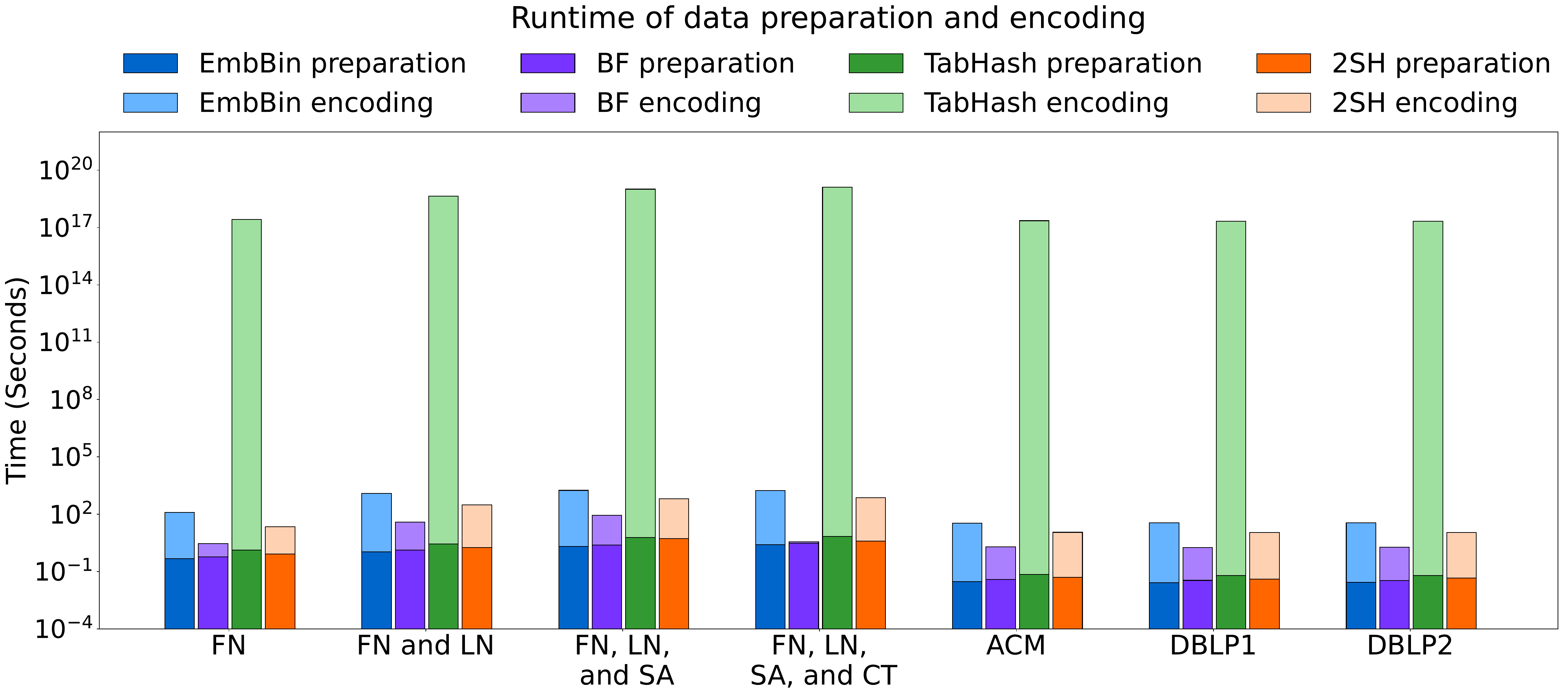}
  \caption{Runtime a DO uses for data preparation and encoding for different approaches on different data sets.}
  \label{fig:time_do}
\end{figure*}

\begin{figure*}[t!]
  \centering
  \includegraphics[width=0.79\textwidth]{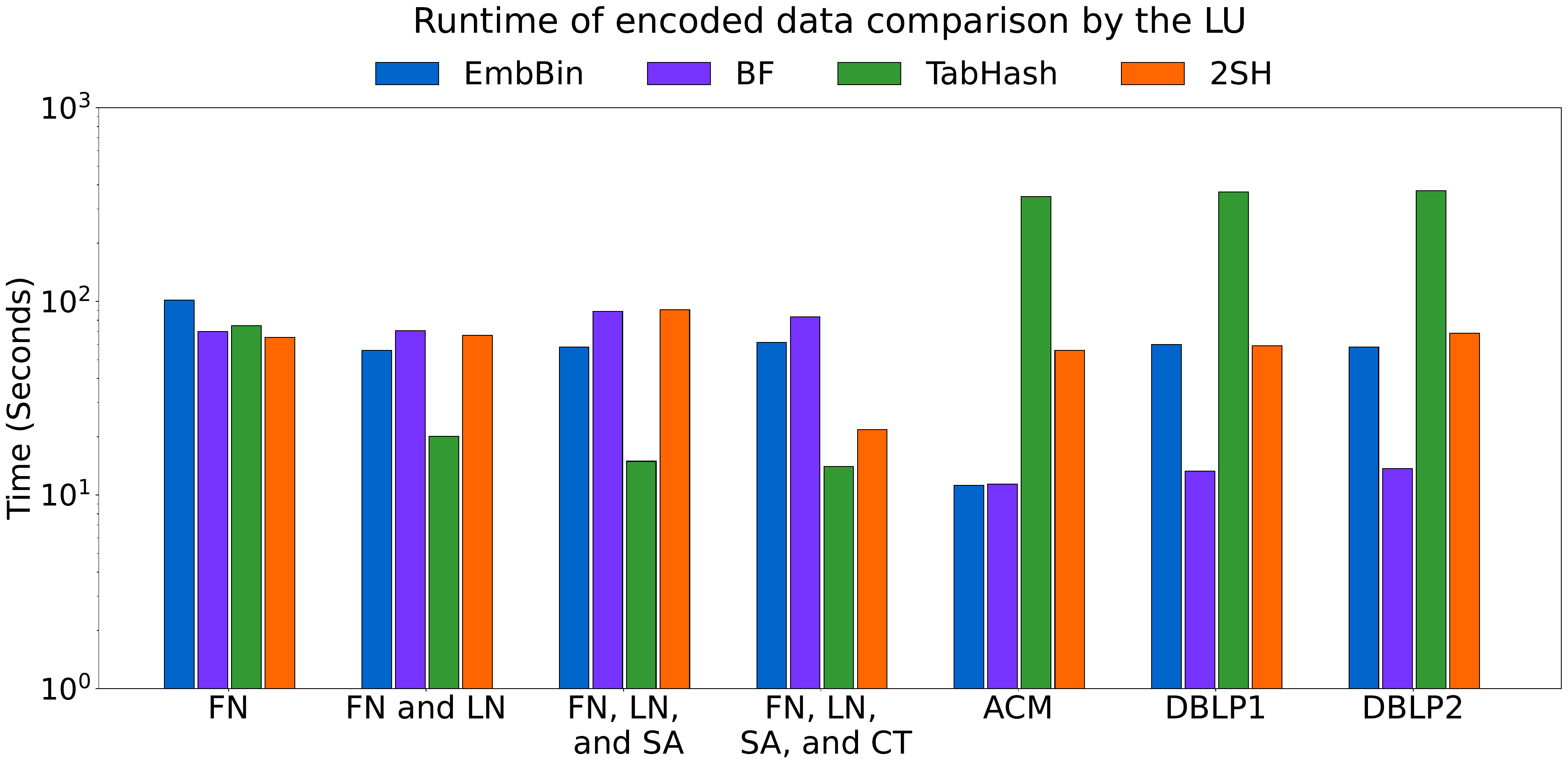}
  \caption{Runtime the LU uses for comparing encoded records for different approaches on different datasets.}
  \label{fig:time_lu}
\end{figure*}

\begin{figure*}[t!]
  \centering
  \includegraphics[width=0.38\textwidth]{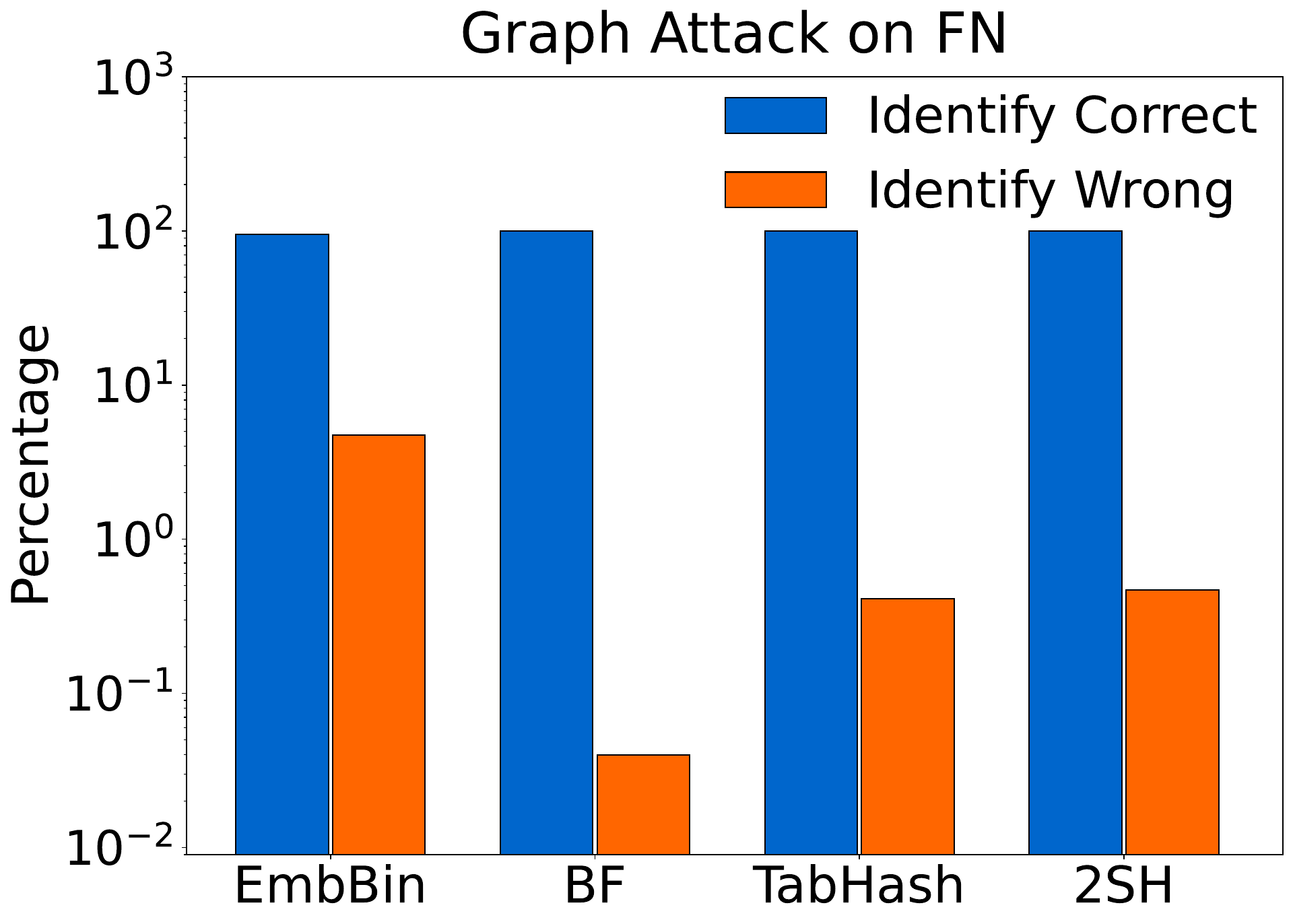} \hspace{5mm}
  \includegraphics[width=0.38\textwidth]{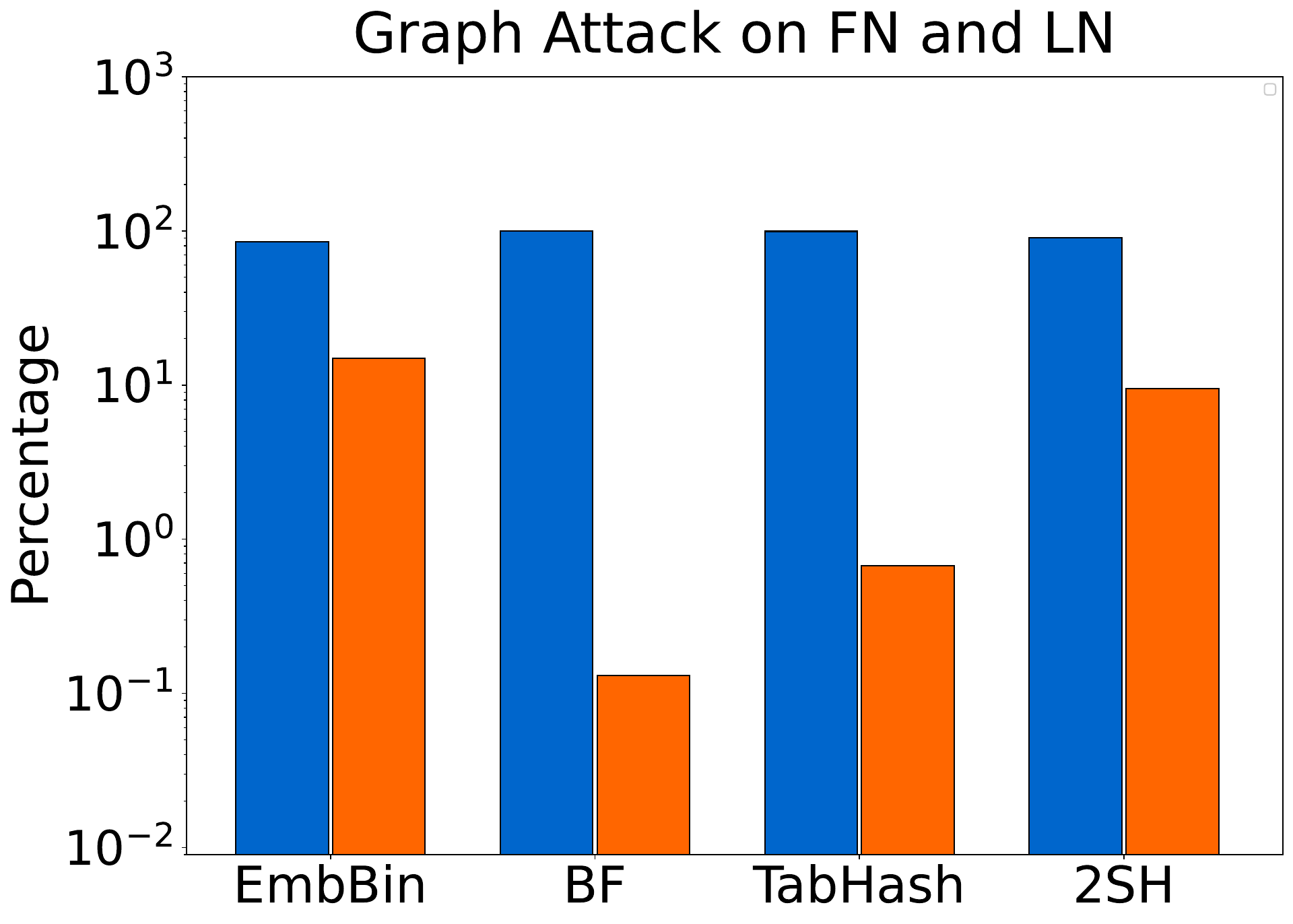}
  \\[3mm]
  \includegraphics[width=0.38\textwidth]{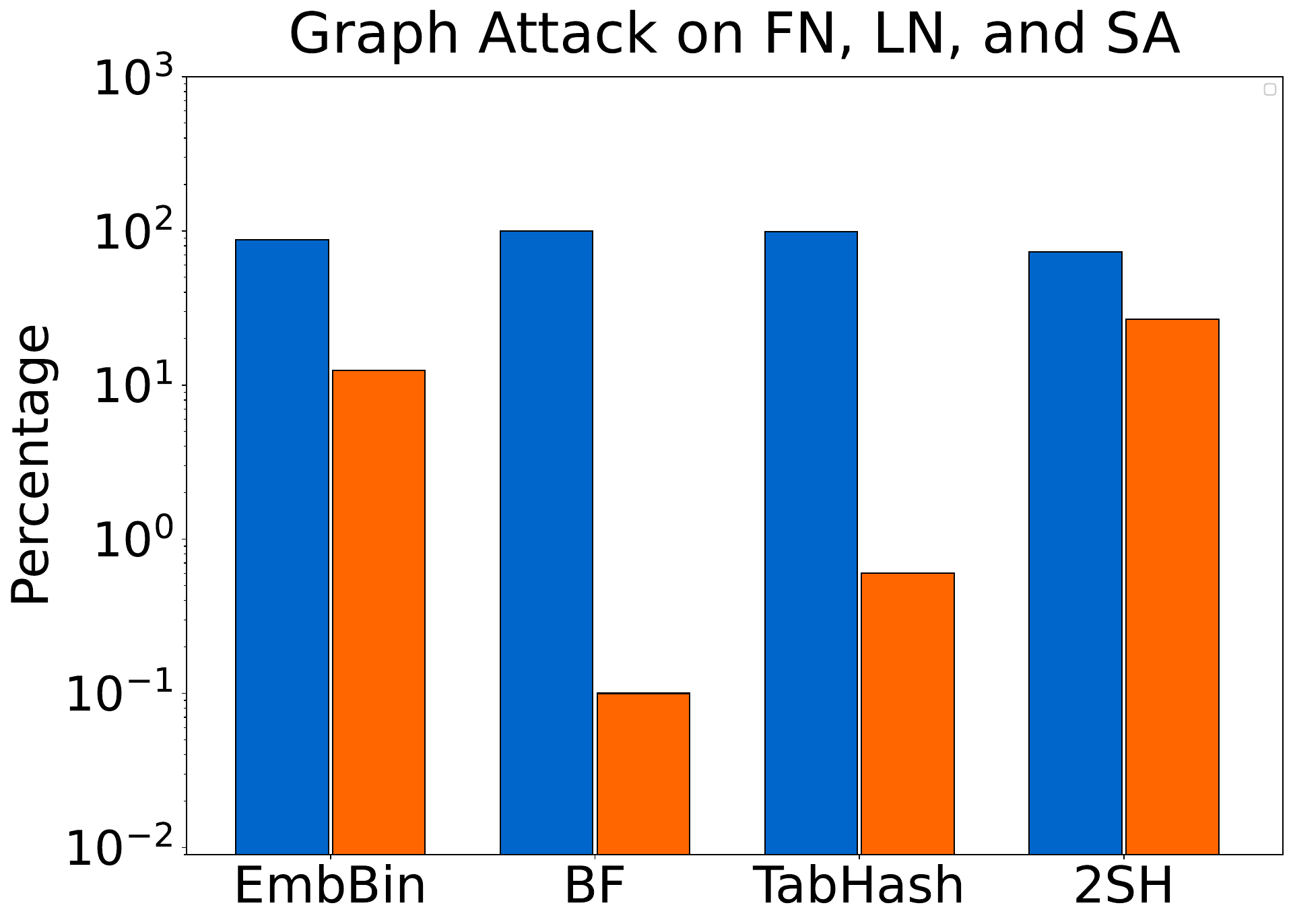}
  \hspace{5mm}
  \includegraphics[width=0.38\textwidth]{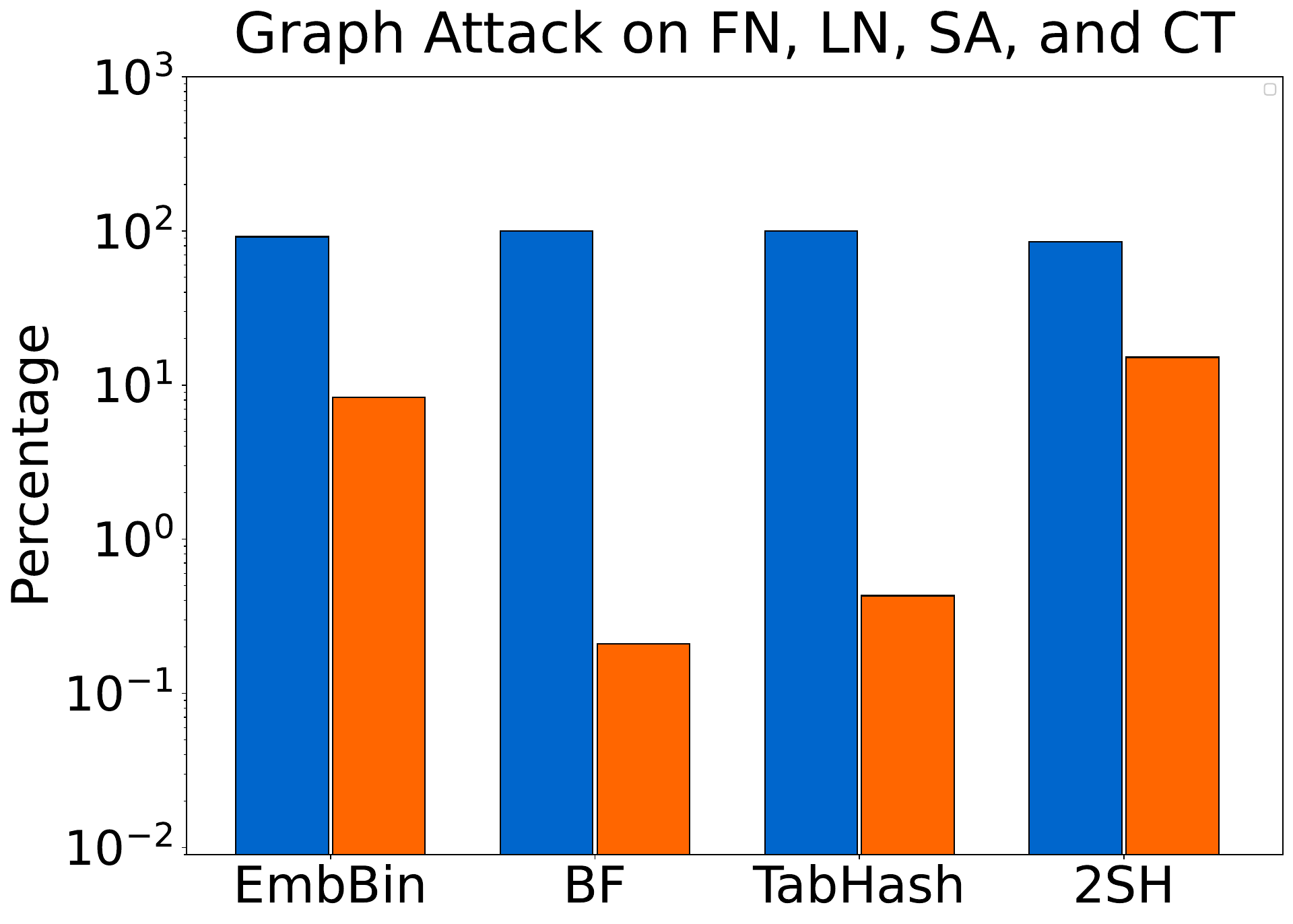}
  \\[3mm]
  \includegraphics[width=0.38\textwidth]{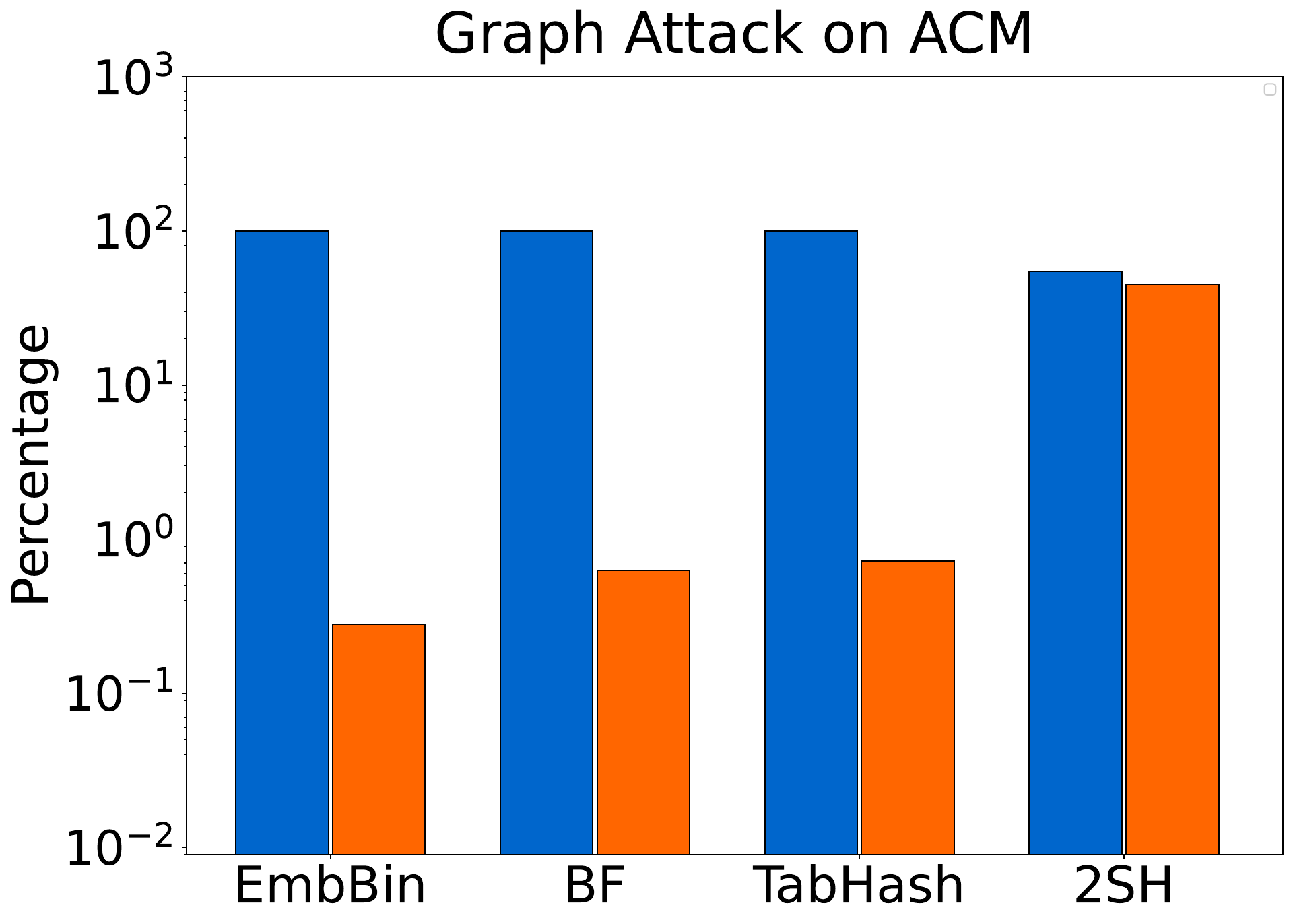}
  \hspace{5mm}
  \includegraphics[width=0.38\textwidth]{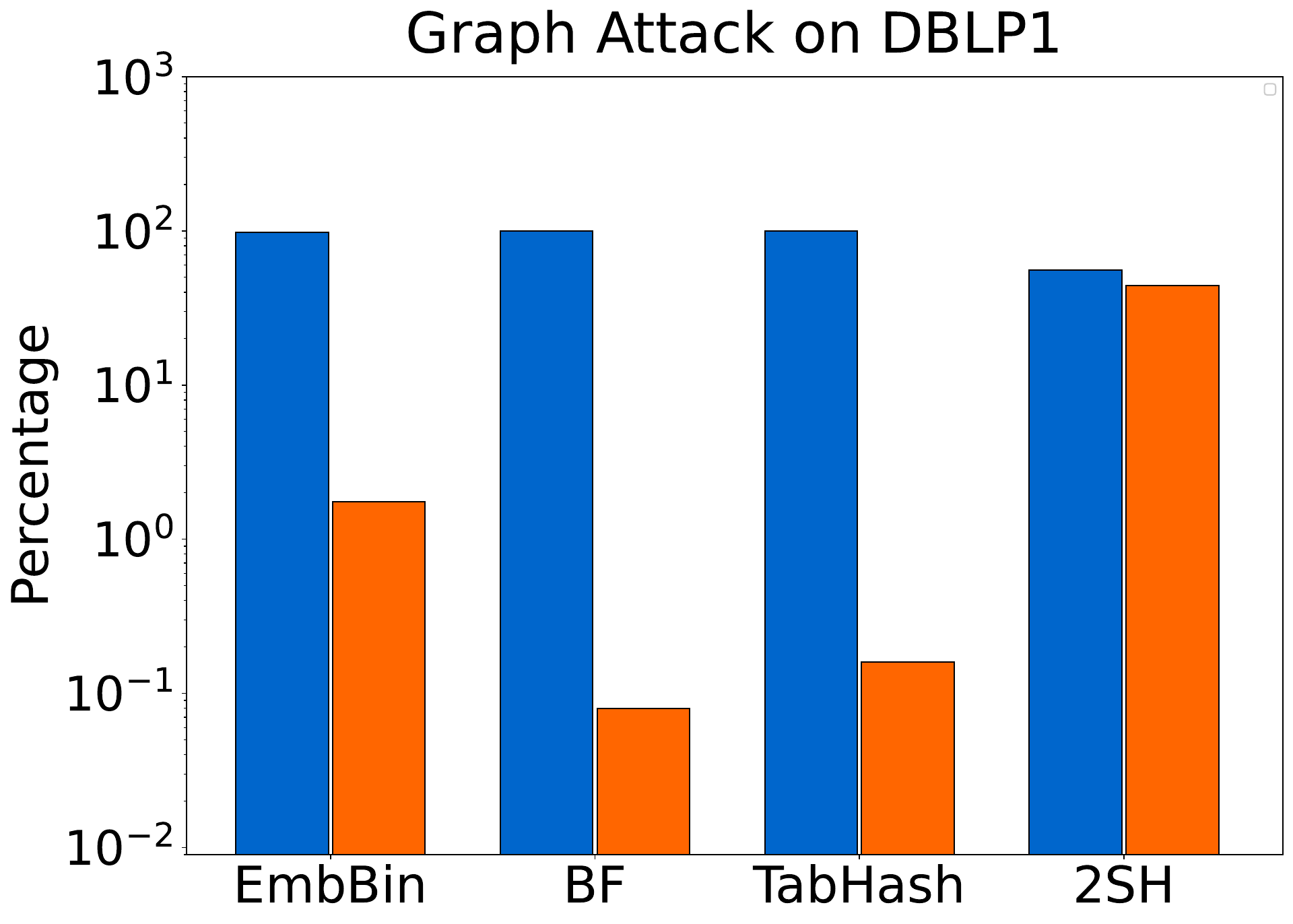}
  \\[3mm]
  \includegraphics[width=0.38\textwidth]{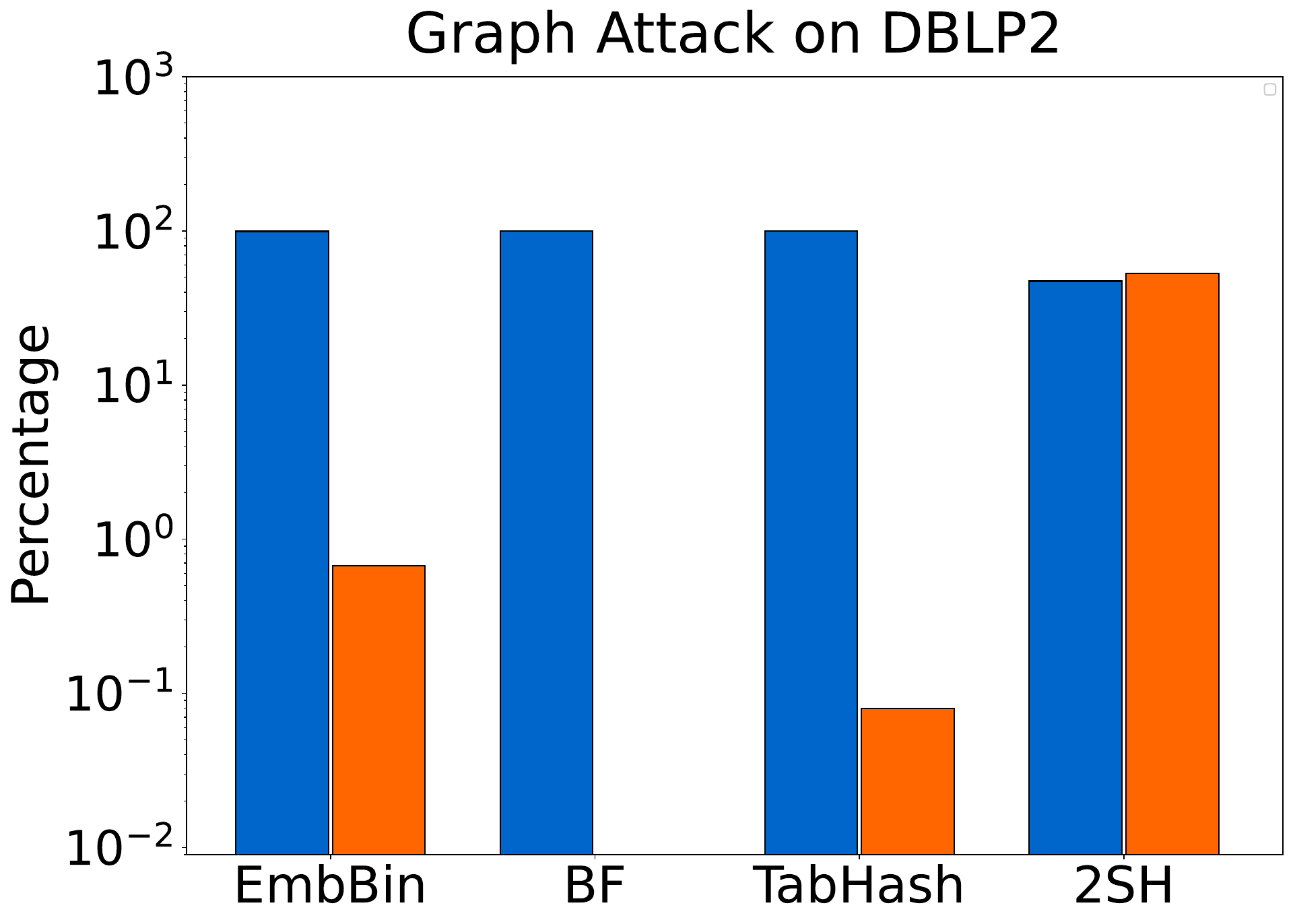}
  \caption{Graph attack on different approaches and data sets.}
  \label{fig:graph_attack}
\end{figure*}

For our approach and the three baselines, we calculated the similarity of plaintext pairs and of encoded pairs using the Dice coefficient similarity~\cite{Chr12}. Depending upon the similarity and the defined similarity threshold $s_t = [0.8,0.9,1.0]$, we find the true positives, false positives, and false negatives, then use them to calculate the precision, recall, accuracy, and F1 measures~\cite{Vat14}. Table~\ref{tab:ncvr_linkquality} and Table~\ref{tab:acmscholar_linkquality} show the measures of the NCVR and DBLP datasets, respectively.

As illustrated in Table~\ref{tab:ncvr_linkquality} and Table~\ref{tab:acmscholar_linkquality}, our approach mostly provides higher than 0.9 for precision, recall, accuracy, and F1. However, for the longer record values such as the FN, LN, SA, and CT, and the ACM datasets, our approach provides low precision values (0.55 for the FN, LN, SA, and CT dataset and 0.57 for the ACM dataset) and F1 (0.73 for the FN, LN, SA, and CT dataset and 0.71 for the ACM dataset). This is because we use $c = mix$ and $l = 2,000$ for these long record values, as they contain both letter and digit values. Then, in the last step of the binarisation, the binary strings of length $l = 2,000$ of these records are randomly selected bits to generate the final binary strings $l_f = 1,000$. Therefore, this can result in lower linkage quality.

Compared to the baselines, our approach outperforms TabHash and 2SH when the record values are shorter, such as for the FN and the FN and LN datasets, while providing similar or worse linkage quality than TabHash and 2SH when the record values are longer, such as for the FN, LN, SA, and CT, and the ACM datasets. The BF mostly outperforms our approach both for short and long record values, except for the DBLP1 dataset pair. Apart from the random bit selection described above, our approach also depends upon the word embedding matrix, which we use CBOW. Therefore, the neighbours of each q-gram can affect to lower linkage quality.


\subsection{Time Complexity Results}
\label{time_complexity_results}

We illustrate the runtimes (in seconds) of the processes by a DO and a LU for each dataset in Figs.~\ref{fig:time_do} and~\ref{fig:time_lu}, respectively. As can be seen in Fig.~\ref{fig:time_do}, our approach and the three baselines consume similar runtimes in the data preparation step. For the data encoding step, our approach uses a similar runtime compared to the 2SH approach, while the TabHash approach uses the longest runtime, but the BF approach uses the shortest runtime. The TabHash uses the longest runtime for encoding because it iteratively selects bits from multiple tabulation hash tables to create the final bit of each position. The BF uses the shortest runtime because it simply encodes q-grams into the bit array, which is less complicated than our approach and the other two approaches.

To reduce time consumption in the comparison process, for all approaches, we applied the phonetic blocking technique~\cite{Chr2020} to the encodings before the encodings are being sent to the LU for conducting comparison. We also limit the number of comparisons to 1 million record pairs because comparing a large number of records will consume much time, such as the NCVR datasets that each contain over 200,000 records.

As can be seen in Fig.~\ref{fig:time_lu}, our approach mostly uses similar runtimes to the BF and the 2SH for the comparison step, except for the ACM dataset which our approach uses less runtime than the 2SH, the FN, LN, SA, and CT dataset which the 2SH uses less runtime than our approach, and the DBLP1 and DBLP2 which our approach uses longer runtime than the BF approach. For the TabHash, it uses the longest runtime in comparison for the ACM, DBLP1, and DBLP2 datasets which generally have very long record values, where it uses the shortest runtimes for the NCVR datasets, except the FN dataset which uses a similar runtime to other approaches.

\subsection{Privacy Results}
\label{privacy_results}

We evaluated the degree of privacy using the graph attack proposed by Vidanage et al.~\cite{Vid20} to reidentify the encoded record values to their corresponding plaintext record values. We assume an adversary uses the same datasets and also the same parameter settings. Fig.~\ref{fig:graph_attack} shows the percentage of correct and wrong reidentification of different approaches on different datasets. A higher percentage of wrong reidentification means an approach is more secure, while a higher percentage of correct reidentification means an approach is less secure.

As can be seen, our approach provides a higher degree of privacy when the record values are shorter, such as the FN, FN and LN, and FN, LN, and SA datasets. Our approach is less secure when the record values in a dataset are very long, such as the ACM, DBLP1, and DBLP2 datasets. However, our approach outperforms the BF and TabHash in all datasets. Compared to the 2SH, our approach provides a higher degree of privacy than the 2SH when the record values are shorter and provides a lower degree of privacy than the 2SH when the record values are longer. In most datasets, the graph attack can reidentify a similar percentage of plaintext record values for the BF and TabHash approaches. Overall, the BF provides the lowest degree of privacy, while the 2SH provides the highest degree of privacy.




\section{Conclusion}
\label{conclusion}

We proposed a privacy-preserving record linkage approach based on embedding and binarisation for linking sensitive data. Overall, our approach outperforms the TabHash in terms of linkage quality, complexity, and privacy, while our approach outperforms the BF in terms of privacy. For the 2SH, our approach provides a higher linkage quality and a higher degree of privacy when the record values are shorter, and consumes similar runtimes both for the processes by a DO and the comparison process by the LU. As future work, we aim to improve linkage quality by using different embedding techniques to generate a word embedding of a record before it is used to generate the binary string.


\section*{Acknowledgments}
This research was funded by King Mongkut's University of Technology North Bangkok, Contract no. KMUTNB-68-NEW-01.


\bibliographystyle{IEEEtran}
\bibliography{citations}

\vfill

\end{document}